\documentclass{aa}
\usepackage{txfonts}
\usepackage{graphicx}
\usepackage{natbib}

\bibpunct{(}{)}{;}{a}{}{,}
\begin{document}

\newcommand{\dd}{\mathrm{d}}
\newcommand{\eps}{\varepsilon}
\newcommand{\msol}{\ensuremath{\mathrm{M}_{\odot}}}
\newcommand{\gccm}{\ensuremath{\frac{\mathrm{g}}{\mathrm{cm}^3}}}
\newcommand{\gcccm}{\ensuremath{{\mathrm{g}}/{\mathrm{cm}^3}}}
\newcommand{\nue}{\ensuremath{\nu_\mathrm{e}}}
\newcommand{\nuae}{\ensuremath{\bar{\nu}_\mathrm{e}}}
\hyphenation{re - pre - sen -ta -tive}
\def\del#1{{}}

\title{On ion-ion correlation effects during stellar core collapse}
\titlerunning{On ion-ion correlation effects during stellar core collapse}
\author{A.~Marek \inst{1}, H.-Th.~Janka \inst{1}, R.~Buras \inst{2}, 
M.~Liebend\"orfer \inst{3}, \and M.~Rampp \inst{1}\thanks{present 
address: Rechenzentrum (RZG) der Max-\-Planck-\-Ge\-sell\-schaft am
Max-Planck-Institut f\"ur Plasmaphysik,
Boltzmannstrasse~2, 85748 Garching, Germany}
}
\authorrunning{A.~Marek et.~al}
\institute{Max-Planck-Institut f\"{u}r Astrophysik,
Karl-Schwarzschild-Str.1, Postfach 1317, 85741 Garching, 
Germany} 
\offprints{H.-Th. Janka,\\ \email{thj@mpa-garching.mpg.de}}
\institute{Max-Planck-Institut f\"ur Astrophysik, 
              Karl-Schwarzschild-Str.\ 1, D-85741 Garching, Germany
                 \and
            Max-Planck-Institut f\"ur Physik (Werner-Heisenberg-Institut),
            F\"ohringer Ring 6, D-80805 M\"unchen, Germany
                 \and
            CITA, University of Toronto, Toronto, Ontario M5S 3H8, Canada}
\date{\today}

\abstract{
The role of ion-ion correlations in suppressing neutrino-nucleus 
elastic scattering during stellar core collapse is reinvestigated,
using two different equations of state. 
We test the improved description by Itoh et al.\ against the
treatment suggested by Horowitz and find that the stronger 
cross section reduction for small momentum transfer in the former 
case does not lead to noticeable changes of the core deleptonization
and entropy increase during collapse, because the improvements are 
relevant below neutrino trapping conditions only for very low
neutrino energies, corresponding to a very small phase space volume.
Treating screening effects for ionic mixtures by the linear mixing 
rule, applied to the collection of representative 
heavy nucleus, $\alpha$ particles, and free nucleons, which is assumed
to characterize the composition in nuclear statistical equilibrium,
we cannot determine mentionable differences during
stellar collapse, because $\alpha$ particles are not sufficiently
abundant and their coherent scattering opacity is too small.
}
\maketitle

\section{Introduction}\label{sec:intro}
Heavy (iron-group or more massive) nuclei dominate the 
composition in stellar iron cores until nuclear densities are
reached in the inner core and the bounce shock raises the 
entropies in the outer core to values where free nucleons 
are favored in nuclear statistical equilibrium (NSE).
During the infall phase therefore coherent,
isoenergetic scattering off nuclei is the
main source of opacity for the electron neutrinos produced by
electron captures \citep[cf., e.g.,][]{bru85,bru89:xp,bru89:eos}. 
Neutrino-nucleus scattering thus hampers the free escape of
neutrinos, is responsible for neutrino trapping around a 
density of $10^{12}\,$g$\,$cm$^{-3}$, and regulates the 
deleptonization and increase of entropy during core collapse.

In the medium of the supernova core nuclei are coupled strongly
with each other by Coulomb forces. They thus form a highly 
correlated
plasma, in which the interactions of neutrinos with 
wavelengths larger than the average ion-ion separation
$a_{\mathrm{ion}}$ (corresponding to neutrino energies 
$\epsilon_\nu \la 2\hbar c/a_{\mathrm{ion}} \sim 20\,$MeV)
are reduced by phase interference effects \citep[][]{itoh75}. The 
corresponding ion screening was more recently calculated
by \cite{hor97} and \cite{Itoh2004}, and investigated in
its effects on stellar core collapse by \cite{brumez97}. The
latter authors employed
the correction factor for neutrino-nucleus
scattering cross sections as given by \cite{hor97}.

\cite{Itoh2004}, however, pointed out that the Monte
Carlo calculations, which Horowitz's fit was based on,
did not allow him to accurately represent the cross section 
reduction for low neutrino energies, i.e., for energies 
$\epsilon_\nu \la \hbar c/a_{\mathrm{ion}}\sim 10\,$MeV, thus
underestimating the importance of ion-ion correlation
effects. \cite{Itoh2004} provided a more accurate analytic 
fitting formula by
using the correct behavior of the liquid structure factor for
small momentum transfer in neutrino-nucleus scattering.

The investigations presented in this work have two goals.
On the one hand we aim at studying the differences for 
stellar core collapse and the formation of the supernova
shock which arise from the improved description of ion 
screening as suggested by \cite{Itoh2004}, compared to a
treatment using the formulae of \cite{hor97}. On the 
other hand we intend to explore the sensitivity of the
evolution to ion screening effects associated with the 
ionic mixture of nuclei and nucleons that are present 
during core collapse. In accordance with the treatment of
NSE in current equations of state (EoSs) for
supernova simulations, we consider the nuclear components
to be free neutrons, free protons, $\alpha$ particles, and
one kind of heavy nucleus which is considered as representative
of the NSE distribution of nuclei beyond $^4$He.
Two different nuclear EoSs with largely different $\alpha$ 
mass fractions during core collapse are employed. The first
EoS (``L\&S''), provided by \cite{latswe91}, is based on a
compressible liquid drop model and uses a Skyrme force for the
nucleon interaction \citep[][]{latpet85}. Our choice of the
compressibility modulus of bulk nuclear matter is 180$\,$MeV,
and the symmetry energy parameter 29.3$\,$MeV,
but the differences in the supernova evolution caused by other
values of the compressibility were shown to be
minor \citep[][]{thobur03,swe94}. The second
EoS used here (``Shen'') is the new relativistic mean field EoS of
\cite{shetok_1_98,shetok98} with a compressibility of nuclear matter
of 281$\,$MeV and a symmetry energy of 36.9$\,$MeV.

The paper is organised as follows: In Sect.~\ref{sect:ion_theory} 
the prescriptions for the ion-ion correlation
factor provided by \cite{hor97} and \cite{Itoh2004} are briefly
summarized,
in Sect.~\ref{sect:numerical_simulations} the input (code, initial
stellar model) in our simulations is described, in 
Sect.~\ref{sect:results} the results are presented, and
in Sect.~\ref{sect:conclusions} we finish with conclusions.

\begin{figure}[tpb!]
\resizebox{\linewidth}{!}{
\includegraphics{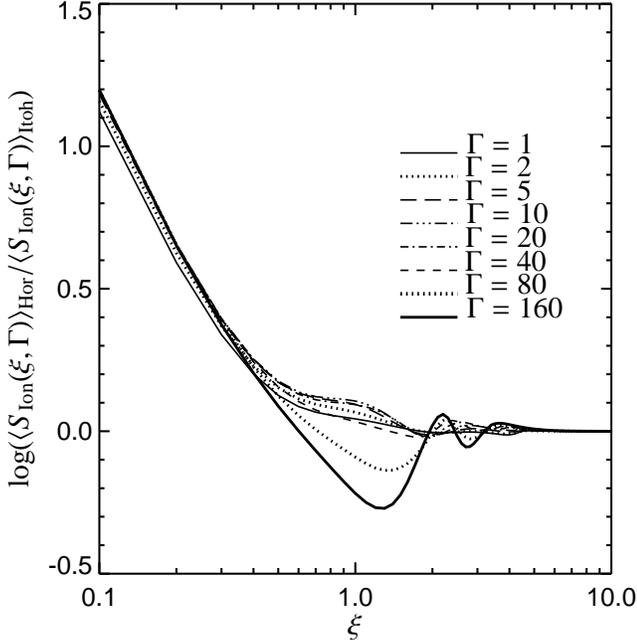}}
\caption{The ratio of the angle-averaged ion-ion correlation factor
as calculated with the fitting formula of \cite{hor97} 
 \citep[see Fig.~9 in][]{brumez97} relative
to the one obtained from the prescription provided by \cite{Itoh2004}
(see their Fig.~2) as a function of the variable $\xi$
for different values of the dimensionless
parameter $\Gamma$.
}\label{fig:comparesfactor}
\end{figure}

\section{Ion-ion correlation factor}\label{sect:ion_theory}

In this study we make use of two different fitting formulae for the 
angle-averaged correlation factor $\left\langle S_\mathrm{ion} 
\right\rangle$, 
which describes the reduction of the neutral-current scattering
of neutrinos off nuclei by ion-ion correlation effects. It is
used as a multiplicative correction to the neutrino-nucleus 
isoenergetic scattering opacity \citep[cf.][]{hor97,brumez97,ramjan02}.
The first formula is provided by \cite{hor97} and is based on Monte 
Carlo results. The second one is given by \cite{Itoh2004} and
was obtained from data calculated with the improved 
hypernetted-chain method 
\citep[see][ and references therein]{Itoh83} for a classical 
one-component plasma. In all simulations with ion screening
we also take into account (the rather small) electron screening 
effects according to \cite{hor97} by applying the additional
correction factor of his Eq.~(19) to the rates of coherent 
neutrino-nucleus scattering.

\begin{table}[tpb!]
\begin{tabular}[width=0.5\linewidth]{|rcll|} \hline
 Reaction & & & References \\ \hline
 $\nu\,e^{\pm}$&$\rightleftharpoons$&$\nu\,e^{\pm}$ &
                                              \cite{mezbru93:nes} \\
             &                    &             &   \cite{cer94}  \\
 $\nu\,A$&$\rightleftharpoons$&$\nu\,A$ & \cite{hor97} \\  
       &                    &       &\cite{brumez97} \\
 $\nu\,N$&$\rightleftharpoons$&$\nu\,N$ & \cite{bursaw98}   \\
 $\nu_e\,n$&$\rightleftharpoons$&$e^-\,p$ & \cite{bursaw99}   \\
 $\bar{\nu}_e\,p$&$\rightleftharpoons$&$e^+\,n$ & \cite{bursaw99}   \\
 $\nu_e\,A'$&$\rightleftharpoons$&$e^-\,A$ &
  \cite{bru85}, \cite{lan2003}\\
    &                    &                  &\cite{mezbru93:coll}   \\
 $\nu\,\bar{\nu}$&$\rightleftharpoons$&$e^-\,e^+ $ & \cite{bru85},
 \cite{ponmir98}   \\
 $\nu\bar{\nu}\,NN$&$\rightleftharpoons$&$NN$ &\cite{hanraf98} \\
 $\nu_{\mu,\tau}\bar{\nu}_{\mu,\tau}$&$\rightleftharpoons$&
$\nu_e\bar{\nu}_e$ &
\cite{burjan03:nunu} \\
$\stackrel{(-)}{\nu}_{\mu,\tau}\stackrel{(-)}{\nu}_e$ &
$ \rightleftharpoons$ & $
 \stackrel{(-)}{\nu}_{\mu,\tau}\stackrel{(-)}{\nu}_e$  &
\cite{burjan03:nunu} \\
\hline
\end{tabular}

\caption[]{Overview of neutrino-matter and neutrino-neutrino
interactions included in our simulations. For each process we provide
reference(s) where more information can be found about physics and
approximations employed in the rate calculations. The numerical
implementation is described in detail in \cite{ramjan02} and 
\cite{burjan03:nunu}. The symbol $\nu$ represents any of the neutrinos
$\nue,\bar{\nu}_e,\nu_\mu,\bar\nu_\mu,\nu_\tau,\bar\nu_\tau$, the
symbols $e^-$, $e^+$, $n$, $p$ and $A$ denote
electrons, positrons, free neutrons and protons, and heavy nuclei,
respectively. The symbol $N$ means neutrons or protons. 
}
\label{tab:reactions}
\end{table}

\subsection{One-component plasma}
\label{sect:one-component}

If the stellar plasma consists of only one nuclear 
species of ions $(Z_\mathrm{ion},A_\mathrm{ion})$, 
the ion sphere radius
which gives the mean interion distance is defined as
\begin{equation}
a_\mathrm{ion}=\left(\frac{3}{4\pi n_\mathrm{ion}}\right)^{1/3} 
\mbox{ ,}
\label{eqn:ionsphere}
\end{equation}
where $n_\mathrm{ion}= n_\mathrm{b} X_\mathrm{ion}/A_\mathrm{ion}$ 
is the number density of the ions with
mass number $A_\mathrm{ion}$, charge $Z_\mathrm{ion}e$, 
and mass fraction $X_\mathrm{ion}$ ($n_\mathrm{b}$ is the
number density of baryons).
The strength of the ion-ion correlations is characterized by 
the dimensionless parameter 
\begin{equation}
\Gamma =
\frac{Z_\mathrm{ion}^2 e^2}{a_\mathrm{ion} k_\mathrm{B} T} 
= 0.2275\, {Z_\mathrm{ion}^2 \over T_{10}} 
\left(\rho_{12}\,{X_\mathrm{ion}
\over A_\mathrm{ion}}\right)^{1/3}
\mbox{ ,}
\label{eqn:gamma}
\end{equation} 
which is the ratio of the unshielded electrostatic potential
energy between the neighbouring ions to the thermal energy. Here
$T_{10}$ denotes the temperature in units of $10^{10}\,$K
and $\rho_{12}$ the mass density in $10^{12}\,$g$\,$cm$^{-3}$.
Note that the definition of $\Gamma$ in \cite{hor97} differs
from the ones used here and in \cite{brumez97} and \cite{Itoh2004}
by a factor $(4\pi)$ in the denominator of Eq.~(\ref{eqn:gamma}).
In Horowitz's notation this factor is absorbed in the employed
value of $e^2$.

\begin{figure*}[tpb!]
\resizebox{\linewidth}{!}{
\includegraphics{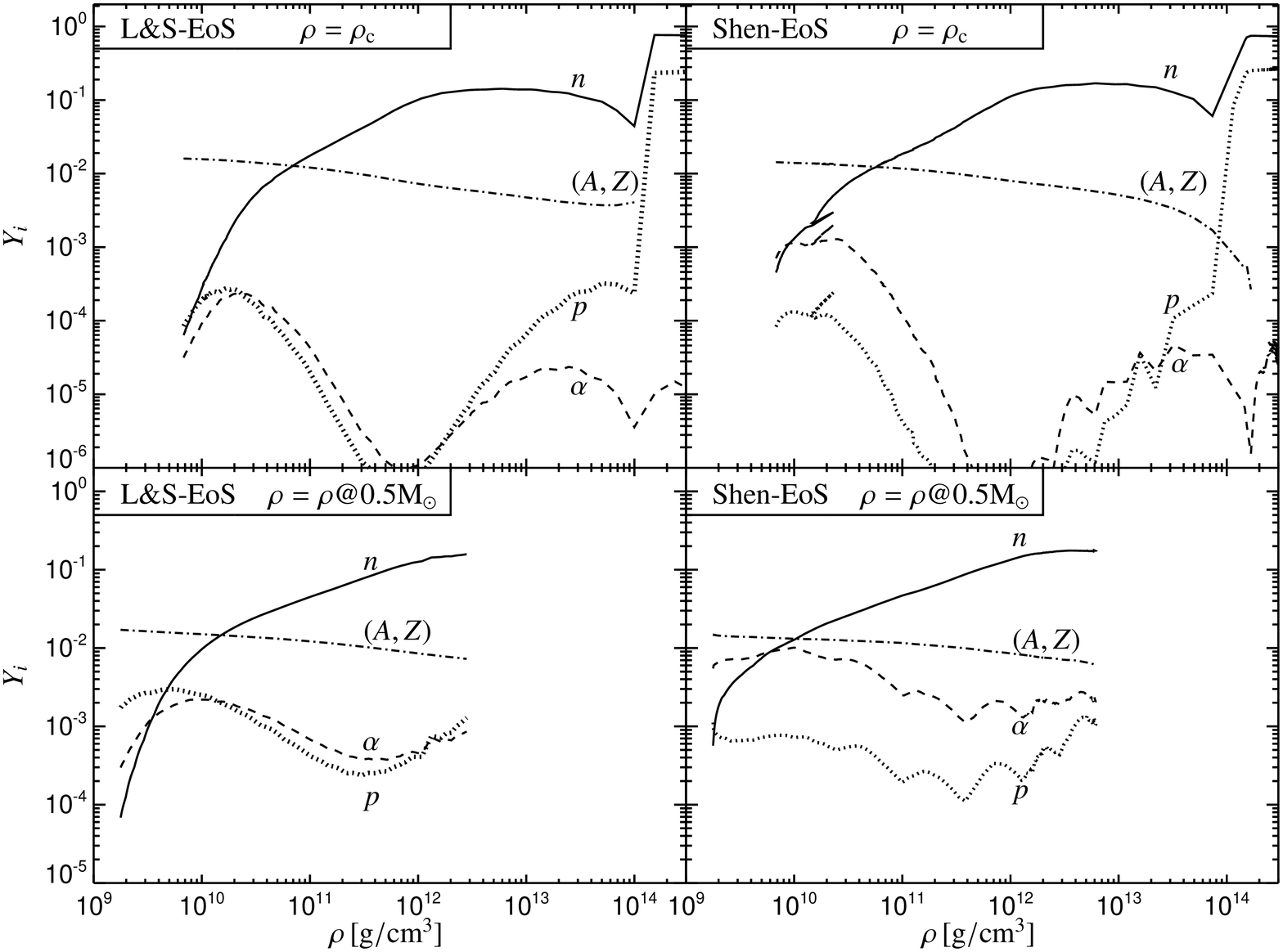}}
\caption{Composition as a function of density at the center
(top) and at an enclosed mass of 0.5$\,M_\odot$ (bottom) for
the evolution during core collapse with the EoSs of Lattimer
\& Swesty (1991; left) and Shen et al.\ (1998a,b; right).
The plots show the evolution of the number fractions $Y$ for 
free neutrons, protons,
$\alpha$-particles, and the representative heavy nucleus until
shock formation. The results were obtained with ion-ion screening
treated according to case ``EoS\_ion\_Itoh'' (see text).}
\label{fig:composition}
\end{figure*}

\cite{hor97} provided the following fitting formula for the 
angle-averaged suppression factor $\langle S_\mathrm{ion}(\xi,\Gamma)
\rangle_\mathrm{Hor}$:
\begin{equation}
\langle S_\mathrm{ion}(\xi,\Gamma) \rangle _\mathrm{Hor} =
\frac{1}{1+\exp \left(- {\displaystyle{\sum_{i=0}^6}}
    \beta_i(\Gamma)\xi \right)} \ ,\ \ \ \mathrm{for} \ \ \
\xi < 3+{4\over \Gamma^{1/2}} \ ,
\label{eqn:shorowitz1}
\end{equation}
and
\begin{equation}
\langle S_\mathrm{ion}(\xi,\Gamma) \rangle _\mathrm{Hor} = 1
\mbox{ ,} \ \ \ \mathrm{otherwise} \mbox{ ,}
\label{eqn:shorowitz2}
\end{equation}
to be applied for $1 \le \Gamma \le 150$; for values of $\Gamma < 1$
or $\Gamma > 150$, \cite{hor97} recommends to simply set $\Gamma$ 
to 1 or 150, respectively. The $\beta_i$ are coefficient functions 
of $\Gamma$ determined from fits to Monte Carlo data.
In Eq.~(\ref{eqn:shorowitz1}) the variable
$\xi$ is the ratio of the mean ion-ion separation $a_\mathrm{ion}$
to the wavelength for neutrinos (during core collapse primarily
electron neutrinos with energy $\epsilon_\nu$), i.e.,
\begin{equation}
\xi=a_\mathrm{ion}\frac{\epsilon_\nu}{\hbar c} \mbox{ .}
\label{eqn:xi}
\end{equation}

\cite{Itoh2004} provide a different fitting formula, see 
Eqs.~(23)--(26) in their paper, which is restricted to the case
of a strongly degenerate electron gas, a usually well fulfilled
condition during stellar core collapse. Their treatment gives
different results for the ion-ion correlation factor, 
$\langle S_\mathrm{ion}(\xi,\Gamma)\rangle$, in the limit
of low neutrino energies ($\xi \la 1$). This can be seen in
Fig.~\ref{fig:comparesfactor}. \cite{Itoh2004} argue that the 
reason for this difference compared to the description by
\cite{hor97} is their correct calculation of the
liquid structure factor $S(k)$ for small momentum transfer $k$.
This makes the suppression of neutrino-nucleus scattering
by ion-ion correlations more important than estimated by
\cite{hor97} and \cite{brumez97}.

The simulations presented here intend to study the 
dynamical consequences of these differences during supernova
core collapse.

Up to now we have considered a stellar medium 
consisting of only one nuclear species and referred to 
formulae derived for a classical one-component plasma. 
However, in the collapsing core of a massive star a mixture
of nuclei besides free neutrons and protons is present, usually
approximated by $\alpha$ particles plus one representative, heavy
nucleus. To deal with that we decided, in the simplest approach,
to calculate the suppression factor
$\langle S_\mathrm{ion}(\xi,\Gamma)\rangle$ for $\alpha$ particles
and for the heavy nucleus independently, using the values 
for the average distance $a_{\mathrm{ion}}$ between ions of the
same kind, derived from the number densities $n_\alpha$ or $n_A$,
respectively. This assumes that different ionic components
coexist without collectively
affecting the screening of neutrino-nucleus interactions. 
Alpha particles thus change the ion screening
for heavy nuclei only by the fact that their presence may reduce 
the number density of heavier nuclei.

\subsection{Ionic mixtures}\label{sect:ionmixture}

For a liquid mixture of different ions $(Z_j,A_j)$ 
including free protons, \cite{Itoh2004} suggest a modified
treatment, referring to earlier work by \cite{itoh79}.
Employing the so-called linear mixing rule, one can extend the
calculations of neutrino-nucleus scattering cross sections,
obtained for a one-component ion liquid, to the case of 
multi-component fluids.

The ion sphere radius for an ion $j$ in the mix is now given by
\begin{equation}
a_j = a_\mathrm{e}\,Z_j^{1/3}
\label{eqn:amixture}
\end{equation}
with $a_\mathrm{e}$ being the electron sphere radius,
\begin{equation}
a_\mathrm{e}=\left(\frac{3}{4\pi \sum_i Z_i n_i}\right)^{1/3} \ ,
\label{eqn:electronsphere}
\end{equation}
where the sum extends over free protons and all nuclei with number
densities $n_i$. The dimensionless variable $\xi_j$ then becomes
\begin{equation}
\xi_j = a_j \,{\epsilon_\nu \over \hbar c} \ .
\label{eqn:ximixture}
\end{equation}
The ion-ion correlations of nuclear species $j$ depend on the
dimensionless parameter $\Gamma_{\! j}$ defined as
\begin{eqnarray}
\Gamma_{\! j}&=&\frac{{Z_j}^{5/3}e^2}{a_\mathrm{e}k_\mathrm{B}T}
= {Z_j^2 e^2\over k_\mathrm{B}T} 
\left( {4\pi\over 3}\sum_i {Z_i\over Z_j}n_i\right)^{1/3} \cr
&=& 0.2275\,{Z_j^{5/3}\over T_{10}} \left(\rho_{12} \sum_i
{X_i Z_i \over A_i} \right)^{1/3}
\mbox{ .}
\label{eqn:averagegamma}
\end{eqnarray}
The angle-averaged ion-ion correlation factor 
$\langle S_{\mathrm{ion}}(\xi_j,\Gamma_{\! j})\rangle$ is now 
evaluated for $\alpha$ particles and heavy nuclei 
with the fitting formula provided by \cite{Itoh2004}, using $\xi_j$
and $\Gamma_{\! j}$ as given in Eqs.~(\ref{eqn:ximixture}) and 
(\ref{eqn:averagegamma}), respectively.

Applying the naive procedure of 
Sect.~\ref{sect:one-component} for the conditions in
a supernova core shows that usually $\xi \gg 1$
for $\alpha$ particles because $\alpha$'s are less abundant
than heavy nuclei in the central part of the core and during 
most phases of the collapse.
Their $\Gamma$ is then less than unity. Both factors
diminish ion screening for $\alpha$'s to a negligible level.
In contrast, following the description in this section,
the presence of $\alpha$'s can affect
also the ion screening of heavy nuclei by reducing
the interion separation (Eq.~\ref{eqn:amixture})
and thus $\xi_j$ (Eq.~\ref{eqn:ximixture}). Moreover, $\Gamma_j$
for heavy nuclei according to Eq.~(\ref{eqn:averagegamma})
might become larger than in Eq.~(\ref{eqn:gamma}).
Therefore the presence of $\alpha$'s has an indirect
influence on neutrino scattering off heavy nuclei and thus 
on ion-ion correlations during stellar core collapse, despite
the fact that the screening effects for $\alpha$ particles
are still small because their $\Gamma$ is usually below 
unity (following \cite{hor97}, $\Gamma$ is then set to unity
for evaluating the angle-averaged cross section suppression
factor).

While our procedure for treating the effects of $\alpha$
particles in an ionic mixture with heavy nuclei and free 
nucleons adopts the recipe of \cite{Itoh2004}, 
\cite{saw05} recently discussed an alternative approach 
to the problem for multi-component fluids by applying 
the Debye-H\"uckel approximation.
He pointed out that in multi-component plasmas the ion-ion
correlation effects might be greatly {\em reduced}, leading
to much larger neutrino opacities than for a one-component
plasma, even if the constituent ions have only a small
range of $N/Z$ ratios. Electron density fluctuations for
an ionic mixture enhance this tendency. Sawyer's calculations
therefore yield a result which is {\em opposite to} our
application of the linear mixing rule for calculating 
multi-component plasma
parameters. They tend to bring one back closer to
the case with ion-ion correlations being ignored, simulations
of which will be presented below, too. Thus we provide a set
of models with the intention to encompass and bracket the 
``extreme'' possibilities discussed in the literature.

\begin{figure}[tpb!]
\resizebox{0.93\linewidth}{!}{
\includegraphics{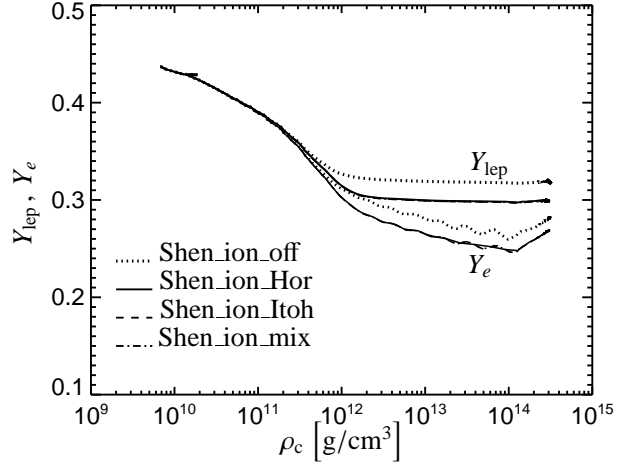}}
\caption{Central electron fraction $Y_e$ and
lepton fraction $Y_\mathrm{lep}$ as functions of
central density for core collapse calculations with the
EoS of \cite{shetok_1_98,shetok98}. Model~Shen\_ion\_off does
not include the effects of ion-ion correlations,
Model~Shen\_ion\_Hor uses the description of ion screening
according to \cite{hor97}, Model~Shen\_ion\_Itoh employs
Itoh et al.'s (2004) treatment for a one-component plasma,
and Model~Shen\_ion\_mix their treatment of ionic mixtures.
}\label{fig:centralye}
\end{figure}

\begin{figure}[tpb!]
\resizebox{0.93\linewidth}{!}{
\includegraphics{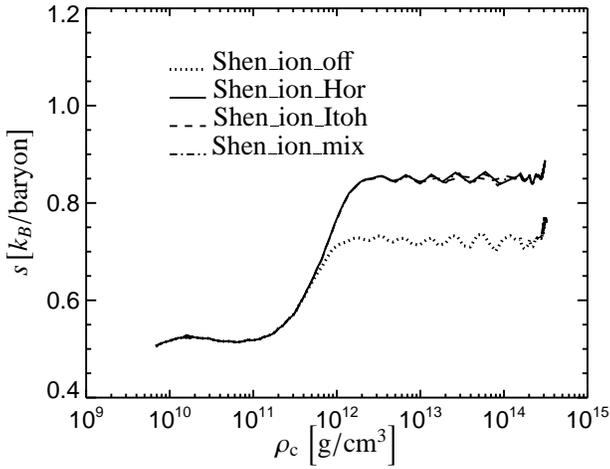}}
\caption{Same as Fig.~\ref{fig:centralye} but for the central
(matter) entropy $s$.
}\label{fig:centrals}
\end{figure}
\begin{figure}[tpb!]
\resizebox{0.94\linewidth}{!}{
\includegraphics{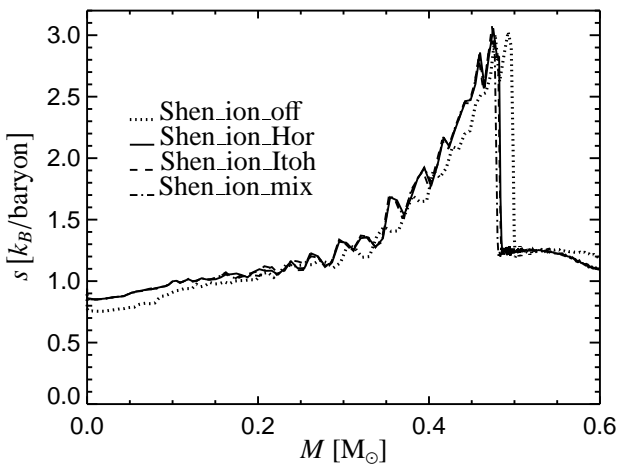}}
\caption{Gas entropy vs.\ enclosed mass at the moment of    
shock formation in the
models of Figs.~\ref{fig:centralye} and \ref{fig:centrals}.  
The shock formation is defined by the instant when the
entropy in the core first reaches a value of 3$\,k_\mathrm{B}$
per nucleon.
}\label{fig:shockformation}
\end{figure}

\section{Numerical simulations}\label{sect:numerical_simulations}

The core collapse simulations performed in this work were
carried out with the
neutrino-hydrodynamics code \textsc{Vertex} in spherical
symmetry. The code is described in detail in \cite{ramjan02}.

\textsc{Vertex} employs the \textsc{Prometheus}
hydrodynamics code \citep[][]{frymue89}, which is a direct
Eulerian implementation of the Piecewise Parabolic Method.
It uses an explicit Godunov type scheme with second order
temporal and third order spatial accuracy, which allows for the
simulation of high Mach number, self-gravitating flows with
discontinuities and can handle complicated equations of state.
The \textsc{Vertex} transport solver employs a
variable Eddington factor algorithm in spherical
coordinates. Angular moments of the O($v/c$) comoving frame
transport equation form a system of lower-dimensional moments
equations, which is closed with a variable
Eddington factor. The latter is determined from a
(simplified) ``model'' Boltzmann equation. General relativistic  
gravity and corresponding effects in the transport are taken
into account approximately. Details about this can be found
in \cite{ramjan02} and \cite{mardim05}. A
comparison with fully relativistic calculations revealed
excellent agreement during core collapse and shock formation 
\citep[][]{lieram05}.

The neutrino interactions used in the simulations of the 
present work are summarized in Table~\ref{tab:reactions}.
Note that neutral-current scatterings of neutrinos off nucleons
and charged-current $\beta$-processes include the effects of
nucleon recoil, thermal motions, and phase space blocking, 
nucleon correlations in dense media \citep[][]{bursaw98,bursaw99}, 
corrections due to the weak magnetism
of nucleons \citep[][]{hor01}, the possible quenching of the
axial-vector coupling in nuclear matter \citep[][]{carpra02}, and
the reduction of the effective nucleon mass at high densities 
\citep[][]{redpra99}. Electron captures on nuclei are
implemented according to the improved treatment
of \cite{lan2003} in regions where NSE holds, taking 
into account the collective
$e$-captures of a large sample of nuclei in NSE with rates
determined from shell model Monte Carlo calculations;
the prescription of \cite{bru85} is used in regions which are
out of NSE. With this
input, the production of $\nu_e$'s by nuclei dominates 
the one by protons during core collapse \citep[][]{lan2003}.

The core collapse simulations presented in this paper were
started from the 15$\,M_{\odot}$ progenitor s15a28 from \cite{wooheg01}.

For describing the thermodynamics and composition of the stellar
plasma, the EoS of \cite{latswe91} and the one of
\cite{shetok_1_98,shetok98} are applied at high densities
($\rho > 6.7\times 10^7\,$g$\,$cm$^{-3}$ 
or $\rho > 2.7\times 10^8\,$g$\,$cm$^{-3}$, respectively).
At lower densities the EoS contains a mixture of electrons,
positrons, photons, nucleons and nuclei, with the nuclear 
composition
being described by a simple approximation to a four-species
NSE for temperatures 
above about 0.5$\,$MeV. Below that temperature the composition
is adopted from the progenitor star and modified if nuclear
burning plays a role during collapse \citep[for details, see 
Appendix~B in][]{ramjan02}. The two EoSs show major differences
in the abundances of $\alpha$ particles, which can be
larger by up to a factor of $\sim\,$10 in case of the 
\cite{shetok_1_98,shetok98}\ EoS. This is visible in
Fig.~\ref{fig:composition} where the number fractions of free 
neutrons, protons, $\alpha$'s and of the representative heavy 
nucleus (whose mass and charge numbers typically grow with
density until nuclei disappear at the phase transition to 
nuclear matter) are displayed as functions of increasing 
density during collapse both at the stellar center and at an 
enclosed mass of 0.5$\,M_{\odot}$. Although their mass fraction
is much lower, $\alpha$ particles in the \cite{shetok_1_98,shetok98} EoS 
can become equally or even more abundant (by a 
factor up to about two) than heavy nuclei in the outer
layers of the collapsing core, in particular exterior to
$M(r)\ga 0.5$--0.6$\,M_{\odot}$.  

For each of the employed EoSs four core collapse simulations were
performed, all starting from the onset of gravitational instability
and carried on until the moment of shock formation. 
Calculations with ion-ion correlation (and electron screening)
effects in neutrino-nucleus scattering being
switched off, i.e.\ for $\langle S_\mathrm{ion} \rangle
\equiv 1$ (and $R_{\mathrm e} \equiv 1$ instead of Eq.~19 of 
Horowitz 1997), 
are denoted with ``EoS\_ion\_off'', where EoS stands
for ``L\&S'' or ``Shen''. They are compared with simulations 
(models ``EoS\_ion\_Hor'') where the ion-ion correlation factor of
Horowitz (1997; Eqs.~\ref{eqn:shorowitz1},\ref{eqn:shorowitz2}) is
used, and with models in which ion-ion
correlations are described according to \cite{Itoh2004} (models
``EoS\_ion\_Itoh''). Finally, the sensitivity of stellar core
collapse to the treatment of ion screening for ionic mixtures
is investigated by simulations 
(models ``EoS\_ion\_mix'') in which the correction
factors $\langle S_\mathrm{ion}(\xi_j,\Gamma_j)\rangle$ are
calculated from Itoh et al.'s (2004) formulae with $\xi_j$ and
$\Gamma_j$ as given in Sect.~\ref{sect:ionmixture}.

\begin{figure*}[tpb!]
\begin{tabular}{cc}
\resizebox{0.48\linewidth}{!}{
\includegraphics{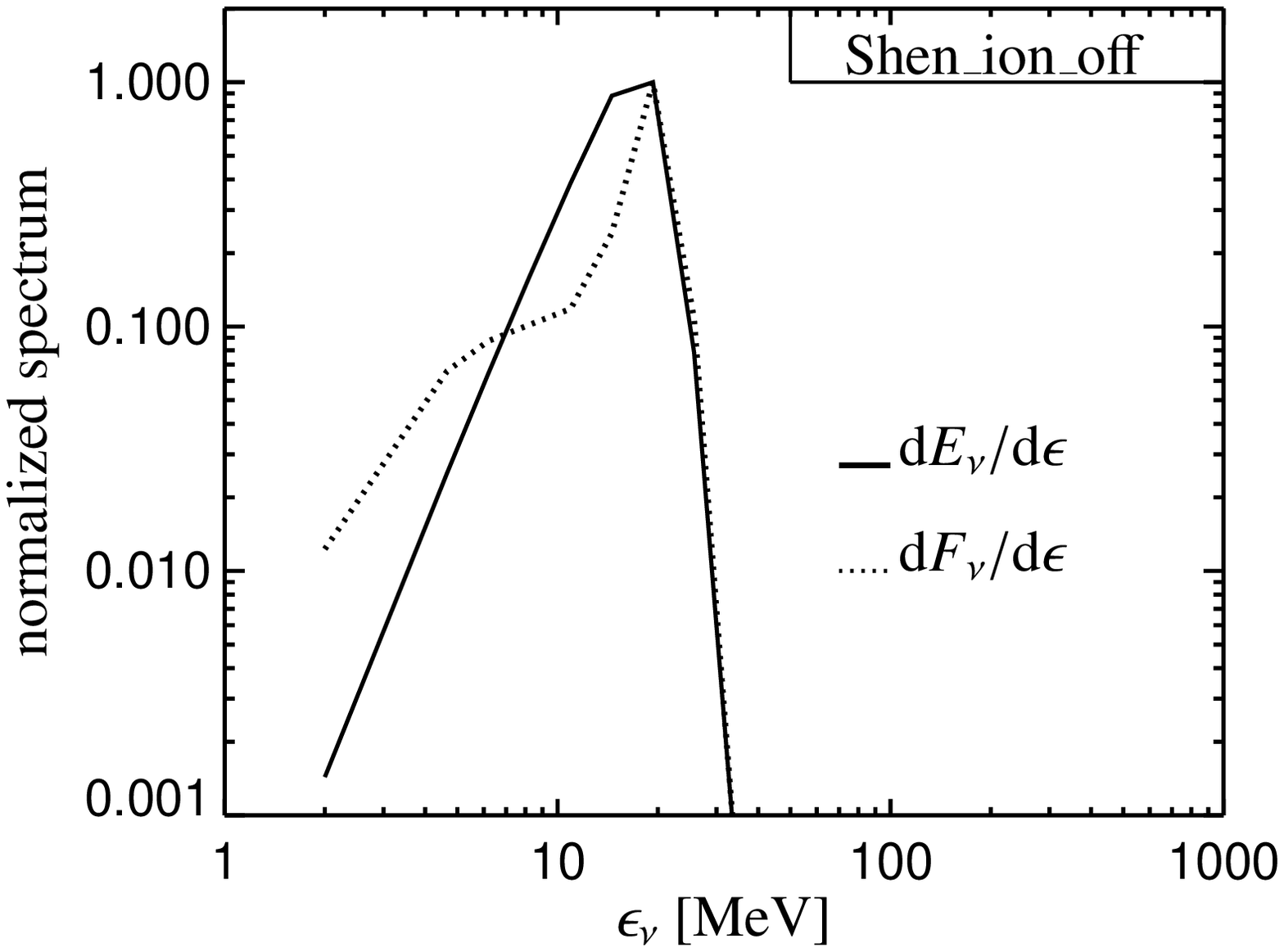}} &  
\resizebox{0.48\linewidth}{!}{
\includegraphics{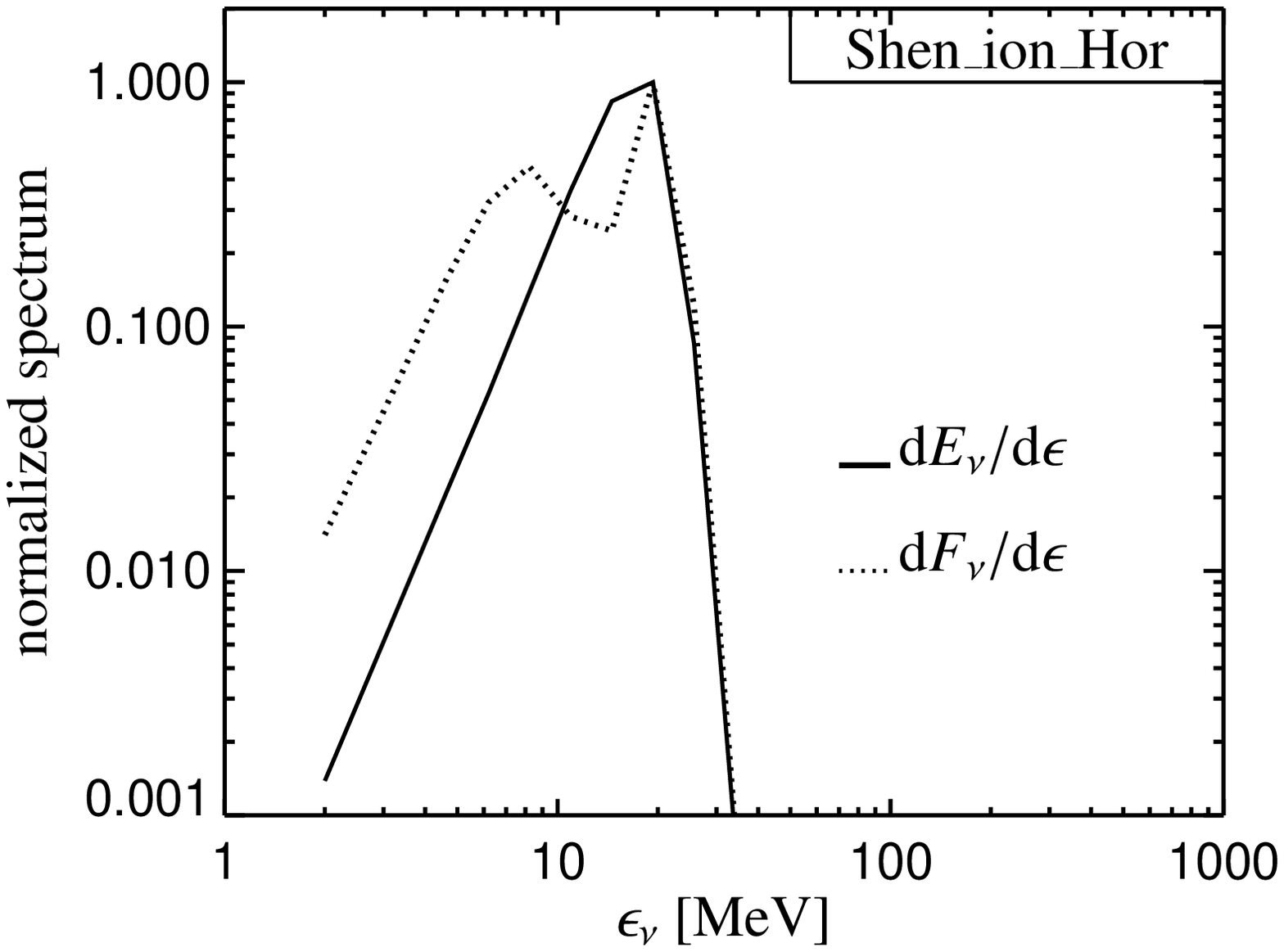}} \\
\end{tabular}
\caption{Spectra of neutrino energy density (solid line)   
and energy flux (dashed)
for Model~Shen\_ion\_off (left) and Model~Shen\_ion\_Hor (right)
when a density of $10^{12}\,$g$\,$cm$^{-3}$ is reached at an  
enclosed mass of 0.3$\,M_\odot$.
}\label{fig:flux+localspectra}
\end{figure*}
\begin{figure*}[tpb!]
\begin{tabular}{cc}
\resizebox{0.48\linewidth}{!}{
\includegraphics{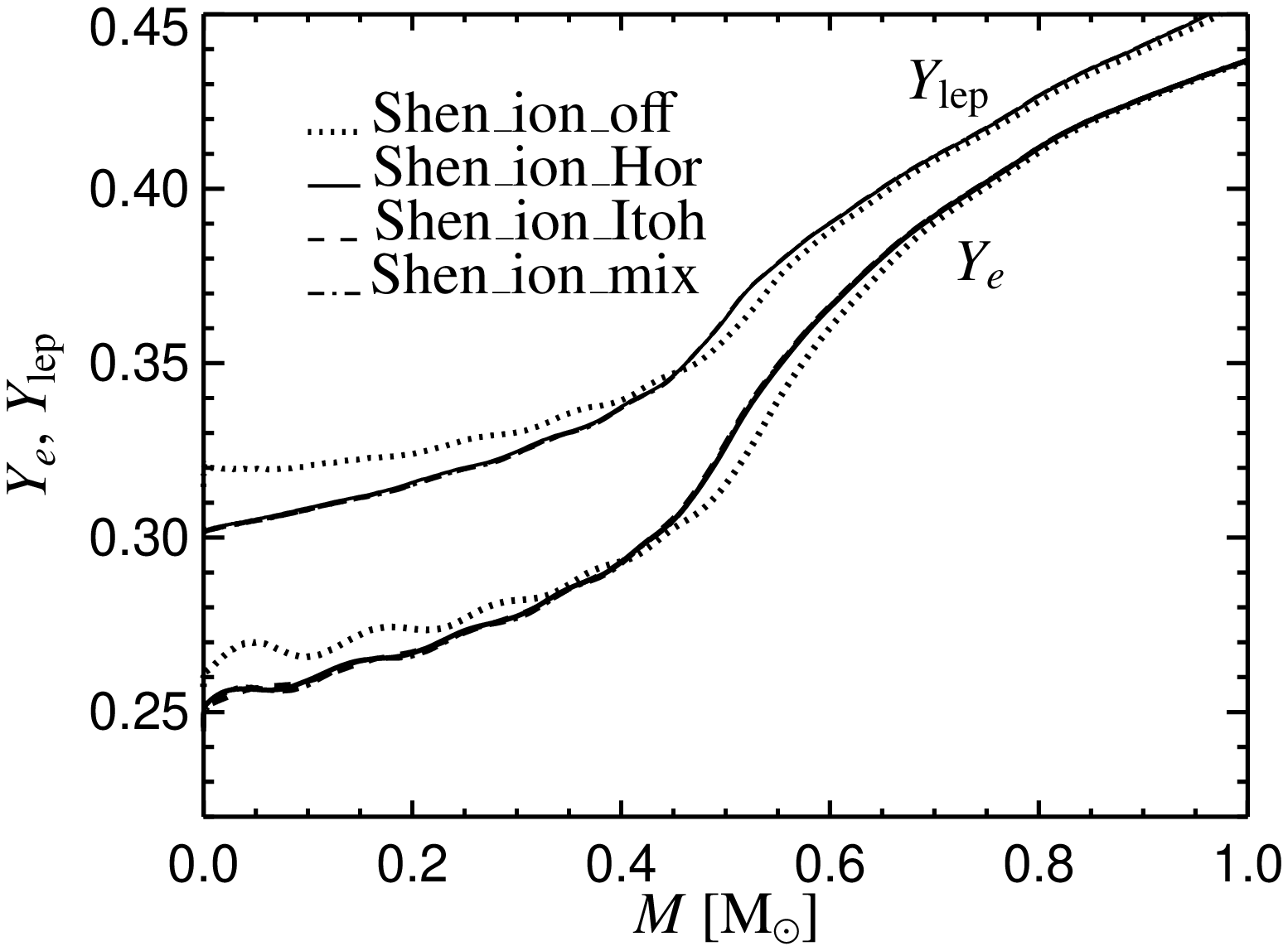}} &
\resizebox{0.48\linewidth}{!}{
\includegraphics{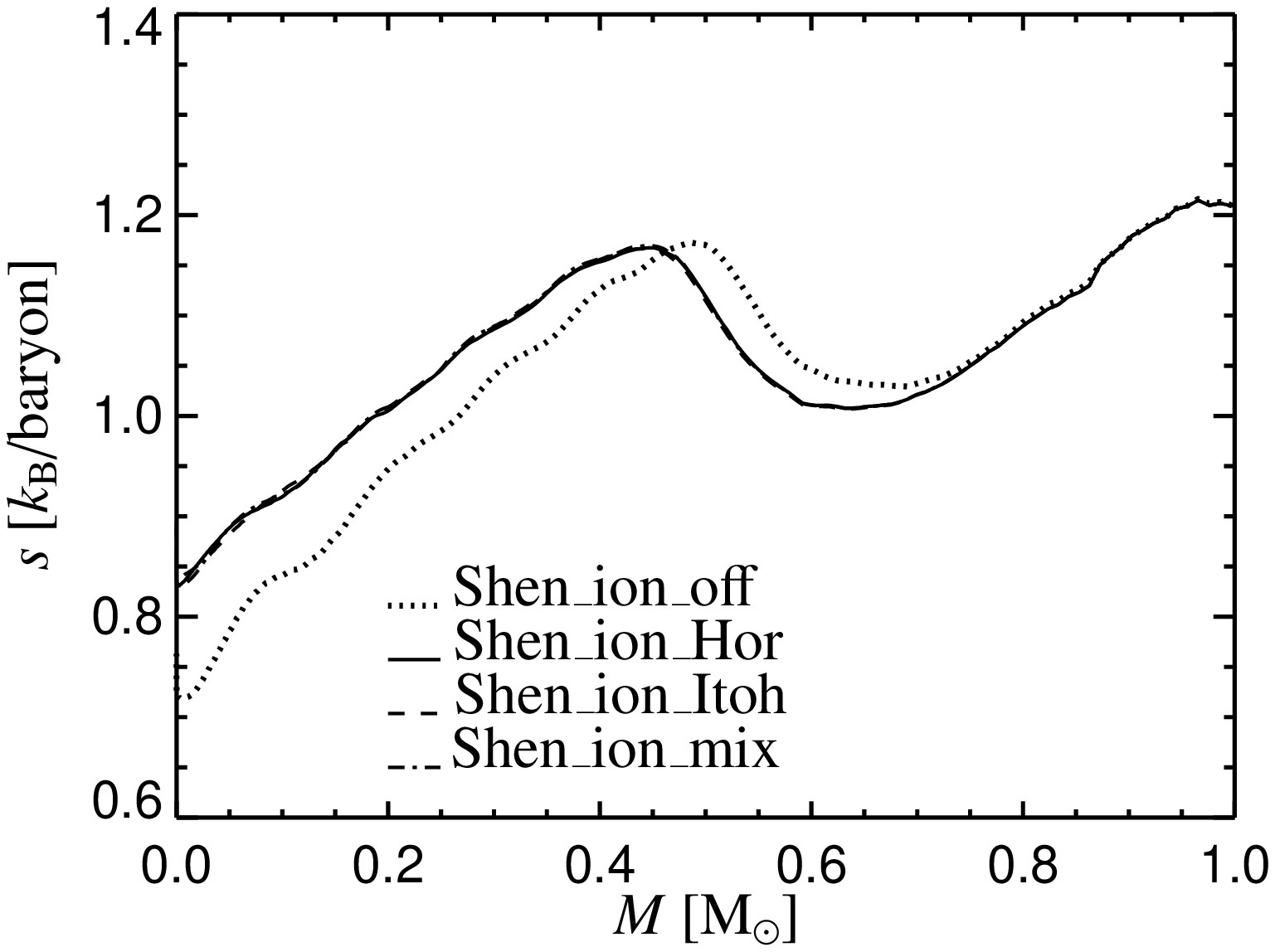}} \\
\end{tabular}
\caption{Profiles of electron fraction $Y_e$, lepton fraction  
$Y_\mathrm{lep}$ (left), and (gas) entropy $s$ (right) versus  
enclosed mass for Models~Shen\_ion\_off (dotted),
Shen\_ion\_Hor (solid), Shen\_ion\_Itoh (dashed), and
Shen\_ion\_mix (dash-dotted) at the time when the central
density has reached a value of $10^{14}\,$g$\,$cm$^{-3}$.
}\label{fig:profiles}
\end{figure*}
\begin{figure*}[tpb!]
\begin{tabular}{cc}
\resizebox{0.48\linewidth}{!}{
\includegraphics{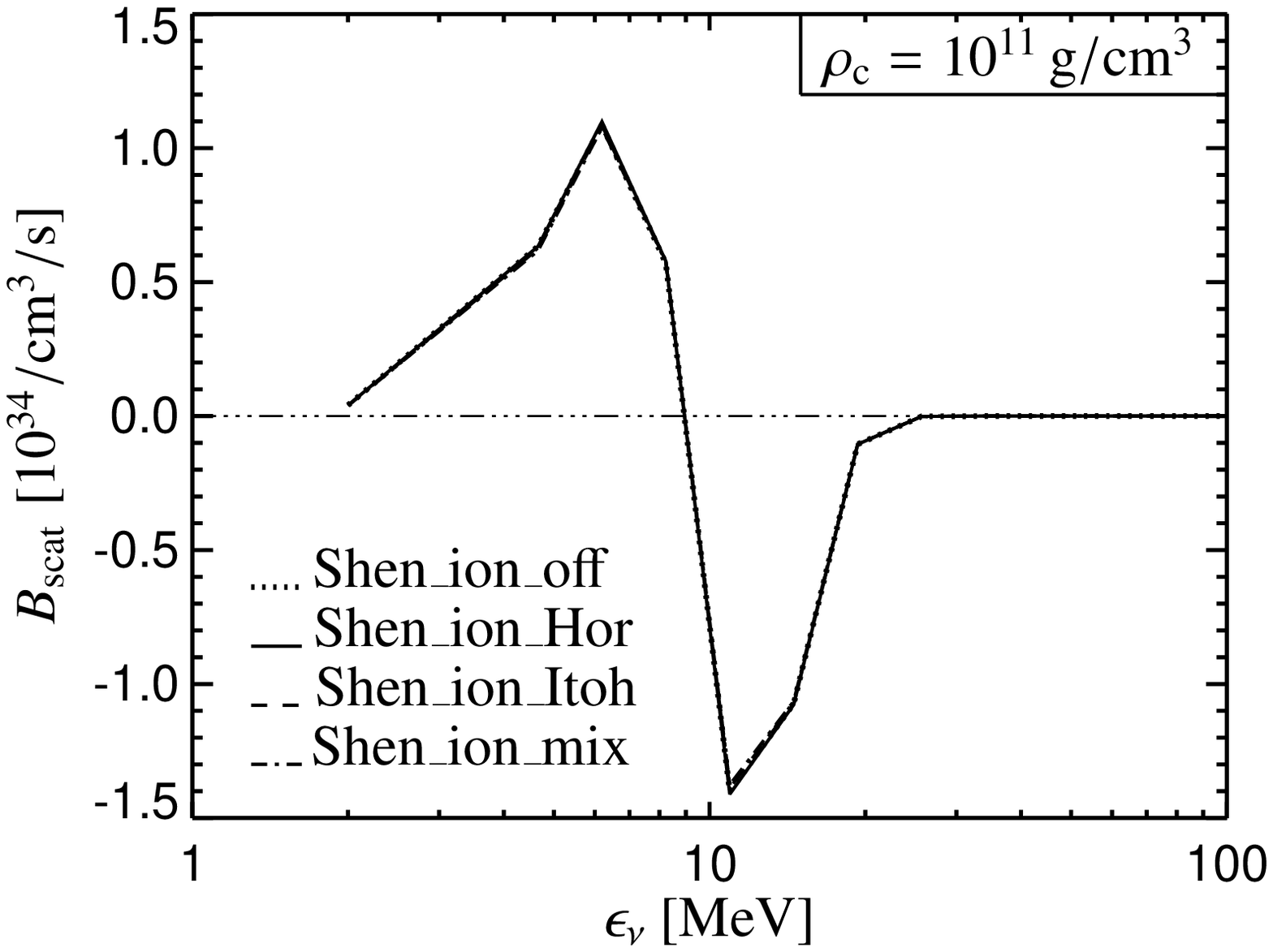}} &
\resizebox{0.48\linewidth}{!}{
\includegraphics{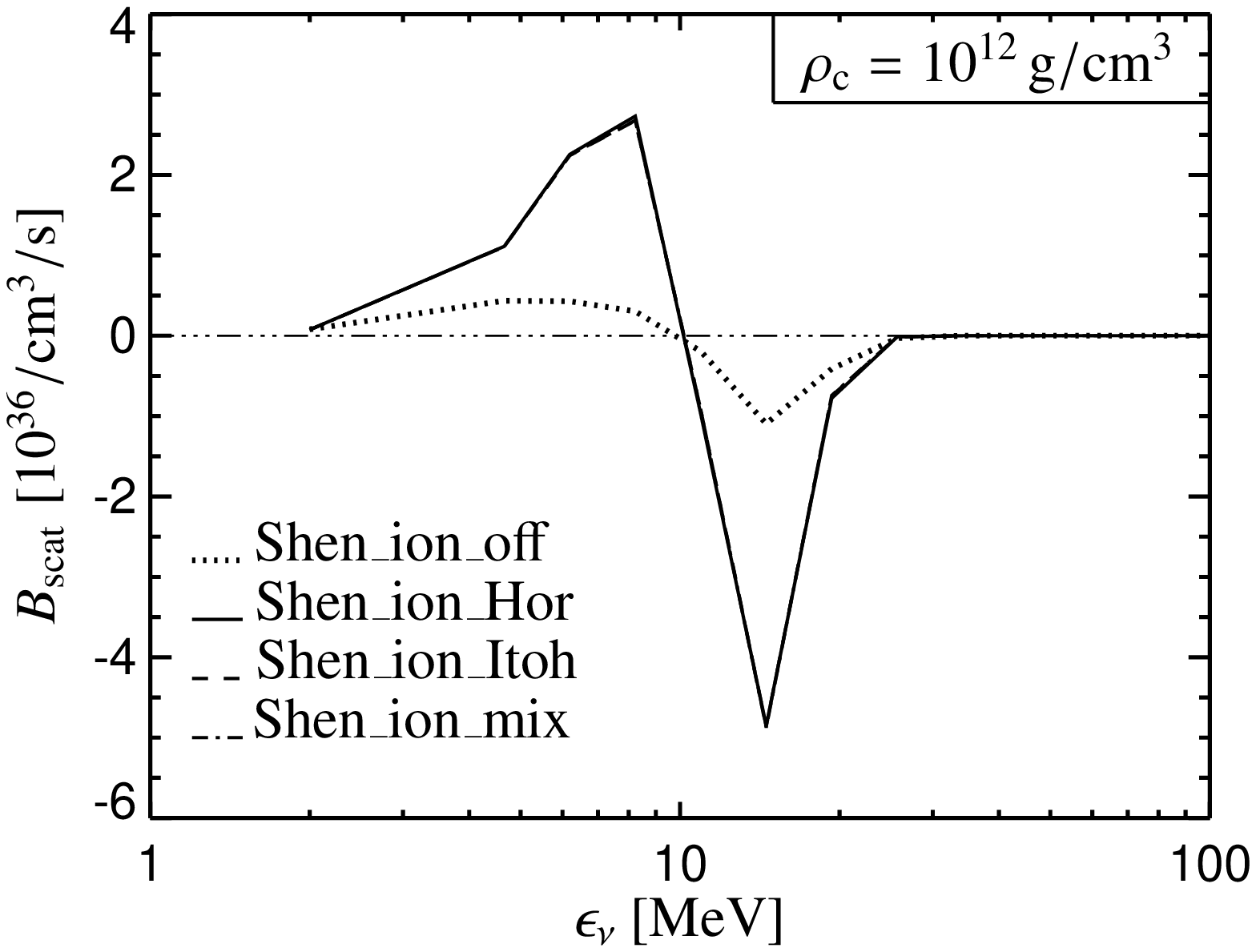}} \\
\end{tabular}
\caption{The energy source term for neutrino-electron
scattering for densities of
(a) $10^{11}\,$g$\,$cm$^{-3}$ and (b) $10^{12}\,$g$\,$cm$^{-3}$
at the stellar center
in the collapse models with the \cite{shetok_1_98,shetok98} EoS.
Negative values mean that neutrino energy is ``absorbed'' (net
scattering out of the corresponding energy bin), positive values
mean ``emission'' of neutrino energy (i.e., net scattering of
neutrinos into the energy bin).}\label{fig:electronscattering}
\end{figure*}

\section{Results}\label{sect:results}

Figures~\ref{fig:centralye} and 
\ref{fig:centrals} show electron fraction $Y_e$, lepton
fraction $Y_\mathrm{lep}$ and (gas) entropy $s$, respectively,
at the core center during collapse simulations with the 
\cite{shetok_1_98,shetok98} EoS. For both EoSs employed in this work,
the same relative changes are found when models without ion 
screening are compared with
calculations with ion-ion correlations according to \cite{hor97}
(see Models Shen\_ion\_Hor and Shen\_ion\_off in
Figs.~\ref{fig:centralye} and \ref{fig:centrals}). We shall mostly
concentrate here on the results obtained with the 
\cite{shetok_1_98,shetok98} EoS, because $\alpha$ particles are 
much more abundant there (see Fig.~\ref{fig:composition})
and many aspects of ion screening in simulations with the
EoS of \cite{latswe91} were already discussed
by \cite{brumez97}. Our results agree qualitatively with those of
the latter paper. Quantitative differences compared to 
\cite{brumez97} are caused by the inclusion of improved electron 
capture rates on nuclei in our work, which significantly increase 
electron captures above a few $10^{10}\,$g$\,$cm$^{-3}$
so that lower values of $Y_\mathrm{lep}$ and $s$ result after 
trapping \citep[cf.][]{lan2003,martinezpinedo05}.

As explained in detail by \cite{brumez97}, the screened cross
section for neutrino-nucleus scattering reduces the transport
optical depth of low-energy neutrinos \citep[cf.\ Fig.~3 in][]{brumez97} 
and allows them to escape from the core more
easily. This is obvious from a 
flux enhancement of neutrinos at energies $\epsilon_\nu
\la 10\,$MeV in Fig.~\ref{fig:flux+localspectra}, where the
situation is displayed at a density of 
$\rho = 10^{12}\,$g$\,$cm$^{-3}$.
Ion-ion correlations thus cause a decrease of $Y_e$ and 
$Y_\mathrm{lep}$ that is stronger by about 0.02 until neutrino
trapping sets in (Fig.~\ref{fig:centralye}).
The homologously collapsing stellar
core correspondingly shrinks and the shock forms at a somewhat 
smaller enclosed mass (Fig.~\ref{fig:shockformation}). 
The shock formation is defined by the moment when the 
postshock entropy first reaches a value of 
3$\,k_\mathrm{B}$ per nucleon. A part of the effect visible in
Fig.~\ref{fig:shockformation} might therefore be a consequence of
the slightly higher core entropy after neutrino trapping in models
with ion screening 
(see Figs.~\ref{fig:centrals} and \ref{fig:profiles}).

\begin{figure*}[tpb!]
\begin{tabular}{cc}
\resizebox{0.5\linewidth}{!}{
\includegraphics{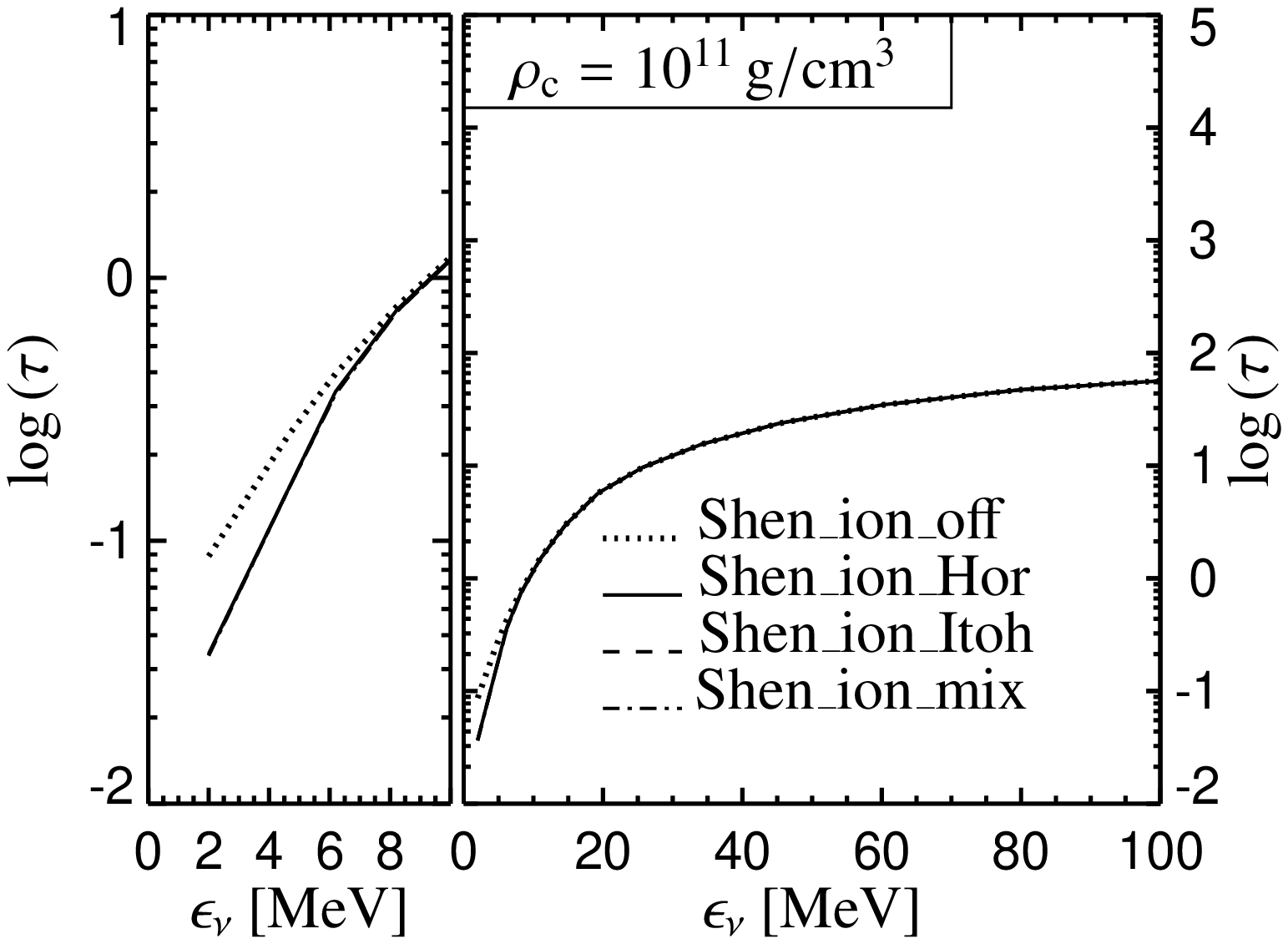}} &
\resizebox{0.5\linewidth}{!}{
\includegraphics{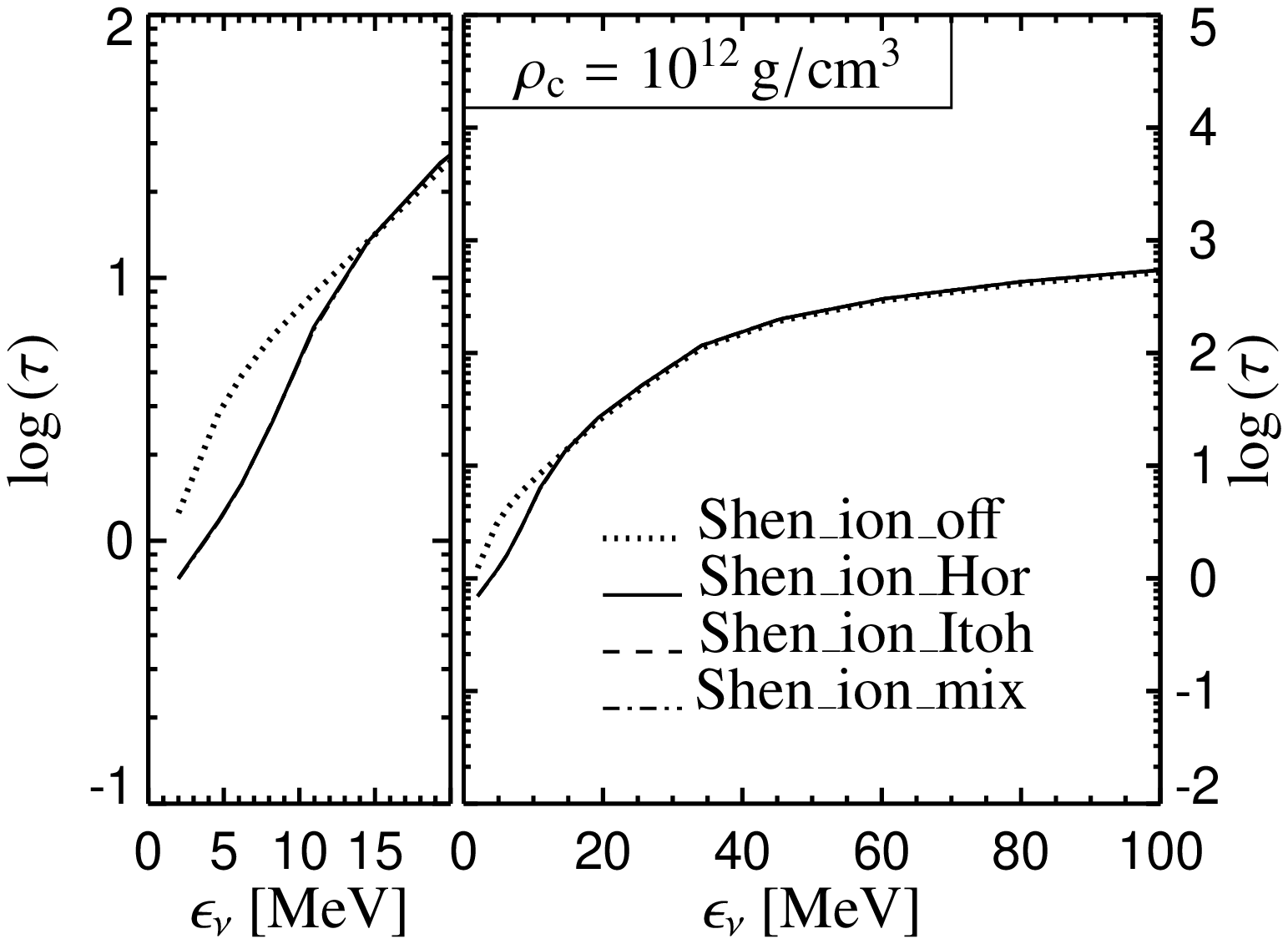}} \\
\bfseries{a} & \bfseries{b} \\
\resizebox{0.5\linewidth}{!}{
\includegraphics{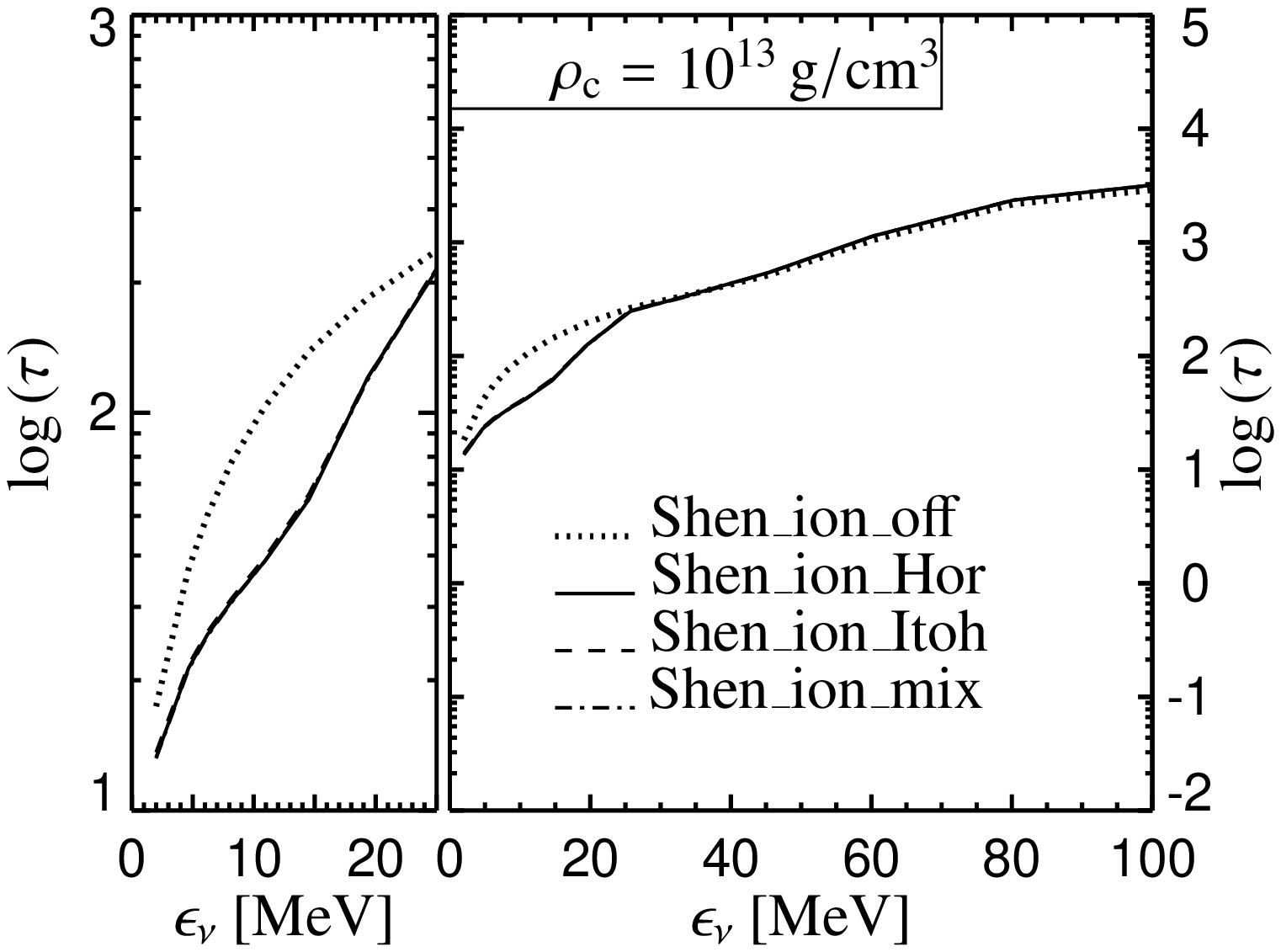}} &
\resizebox{0.5\linewidth}{!}{
\includegraphics{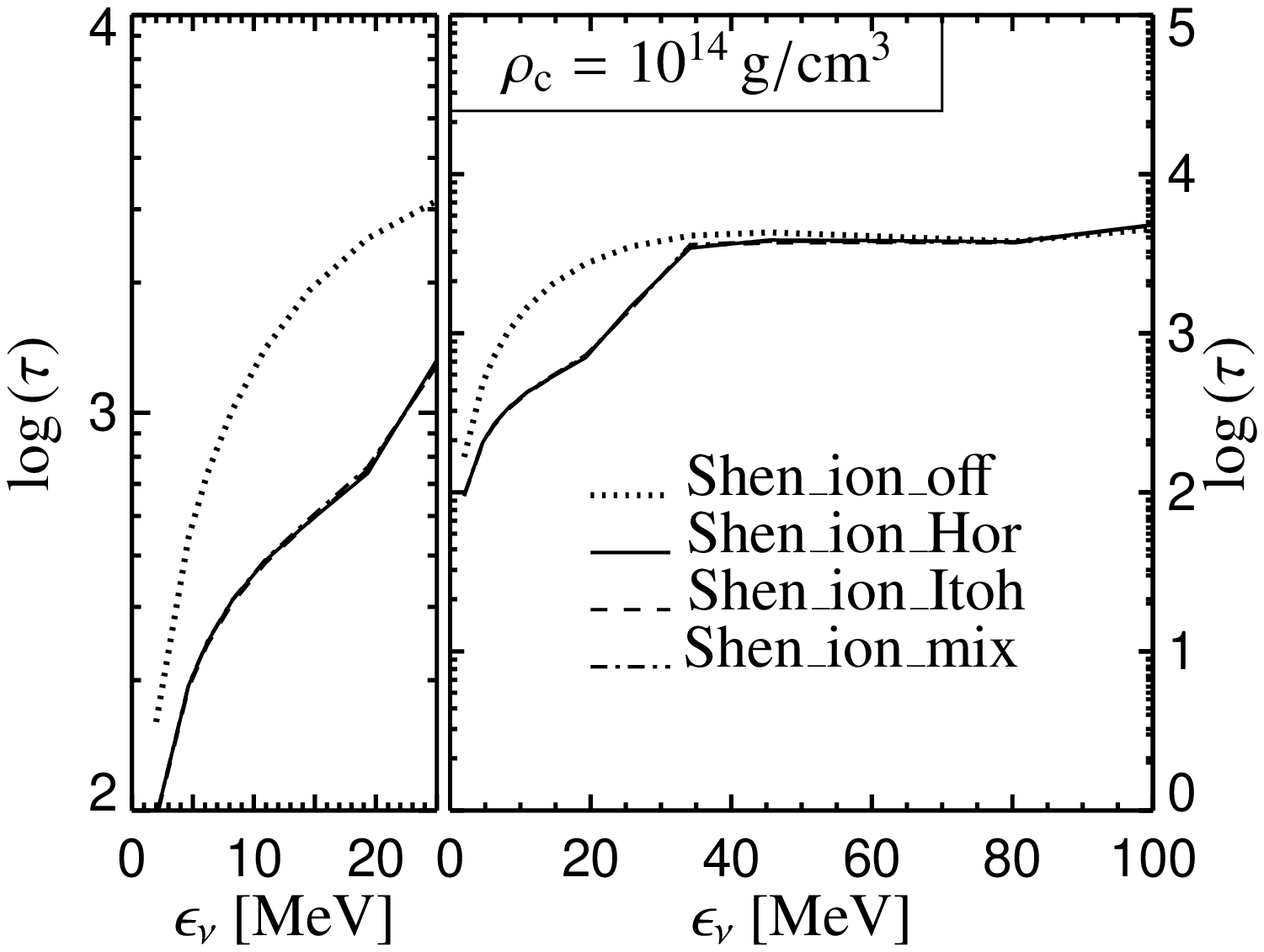}} \\
\bfseries{c} & \bfseries{d} \\
\end{tabular}
\caption{The optical depth for energy exchange between neutrinos
and stellar plasma as a function of the neutrino energy at the
center of the iron core for densities of
(a) $10^{11}\,$g$\,$cm$^{-3}$, (b) $10^{12}\,$g$\,$cm$^{-3}$, (c)
$10^{13}\,$g$\,$cm$^{-3}$, and (d) $10^{14}\,$g$\,$cm$^{-3}$.
The results are taken from collapse calculations with the 
\cite{shetok_1_98,shetok98} EoS.
The left panels show enlargements of the low-energy window where
ion-ion correlations have the largest effect.
}\label{fig:opticaldepth}
\end{figure*}


Neutrino-electron scattering is very efficient
in downscattering neutrinos from the high energies, where they are
mostly created by electron captures, to lower energy states.
Therefore the phase space at low energies is quickly refilled.
Figure~\ref{fig:electronscattering} shows the source term for 
energy redistribution by neutrino scatterings off electrons for
two density values below trapping conditions. The downscattering
of high-energy neutrinos explains why the local energy
spectra, $\mathrm{d}E_\nu/\mathrm{d}\epsilon_\nu$
with $E_\nu$ being the neutrino energy density,
are essentially the same in Models~Shen\_ion\_off and
Shen\_ion\_Hor, despite of clear differences between the energy
flux spectra of both runs (Fig.~\ref{fig:flux+localspectra}).

\begin{figure*}[tpb!]
\begin{tabular}{cc}
\resizebox{0.5\linewidth}{!}{
\includegraphics{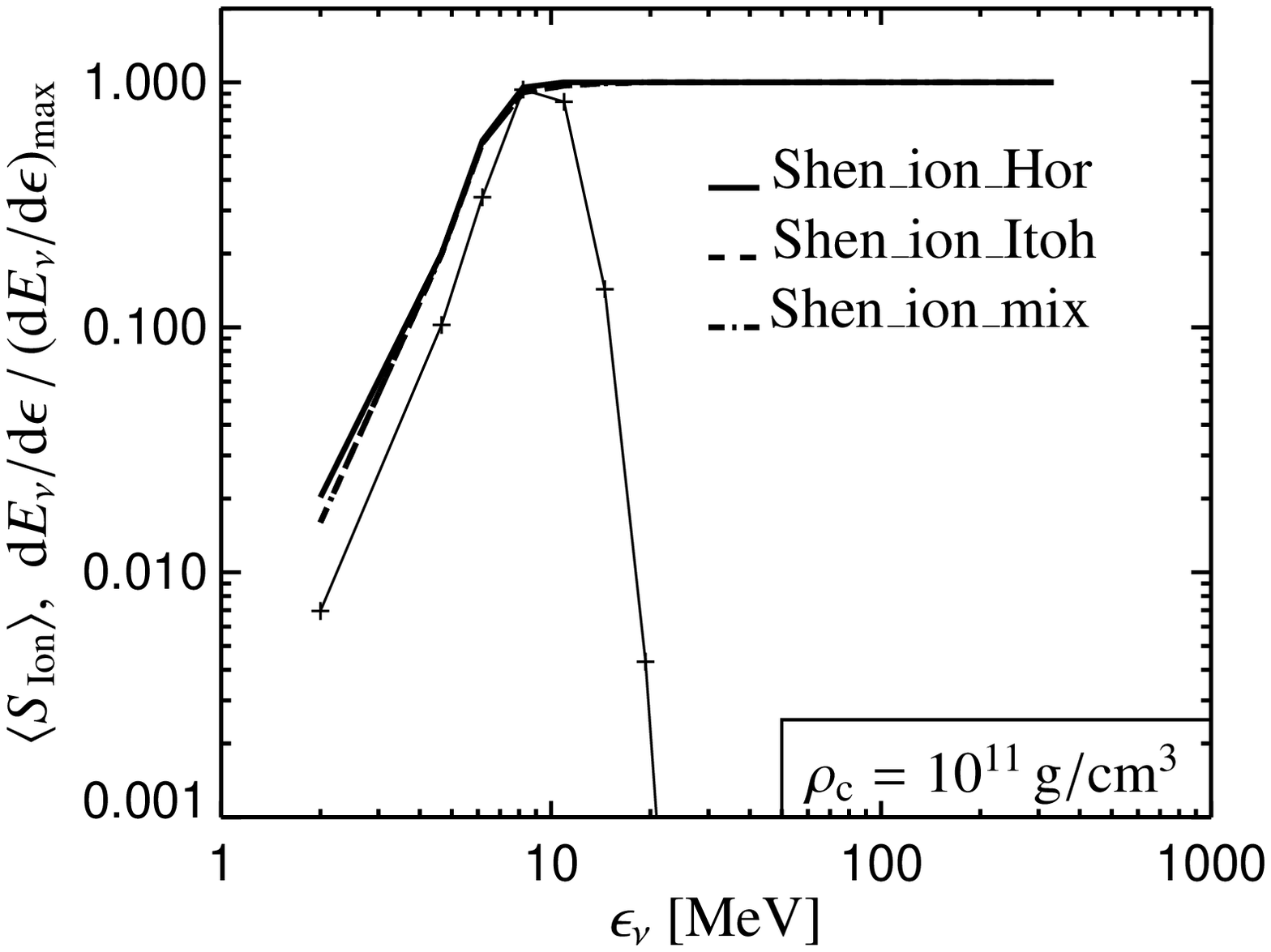}} &
\resizebox{0.5\linewidth}{!}{
\includegraphics{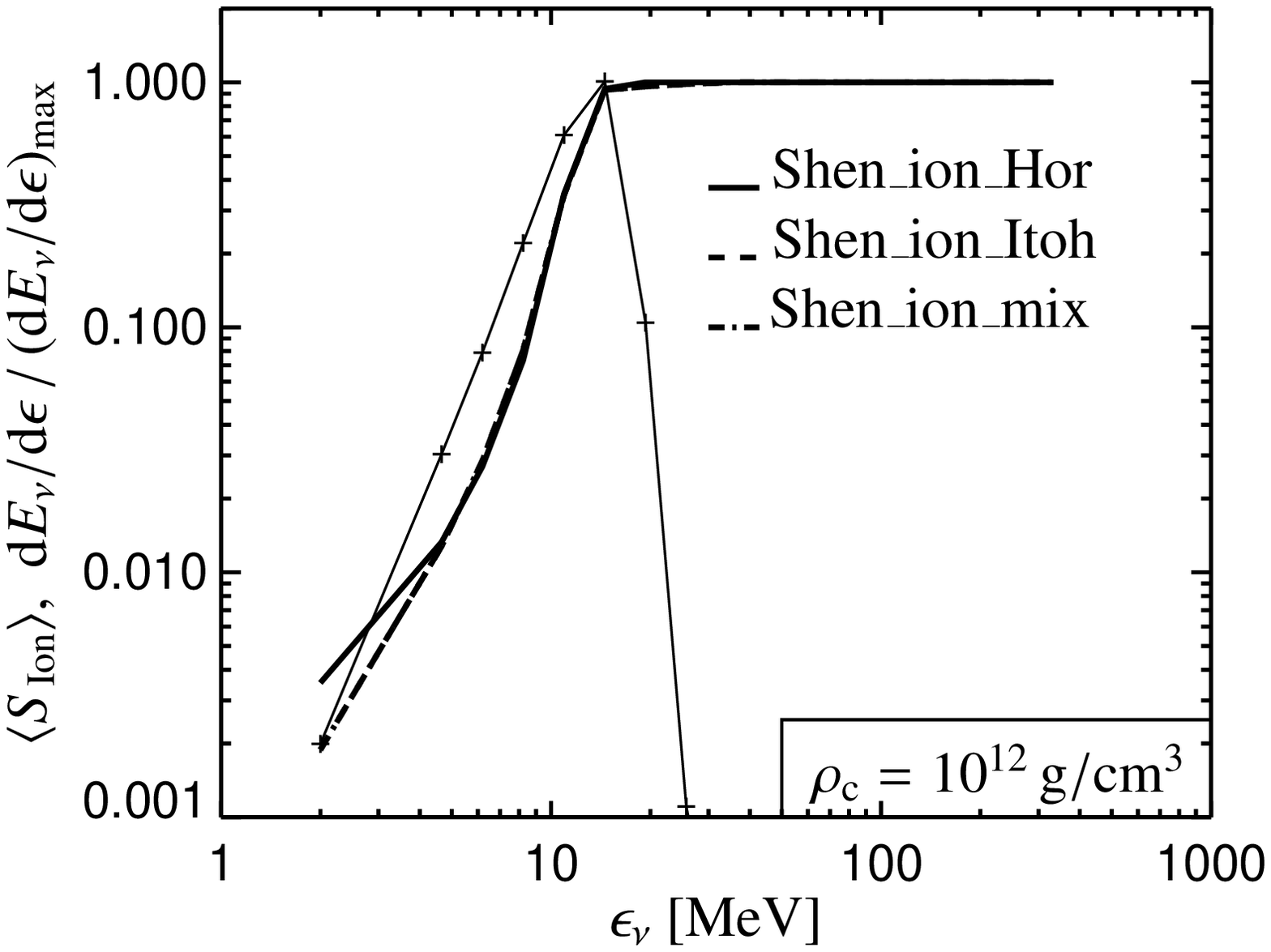}} \\
\bfseries{a} & \bfseries{b} \\
\resizebox{0.5\linewidth}{!}{
\includegraphics{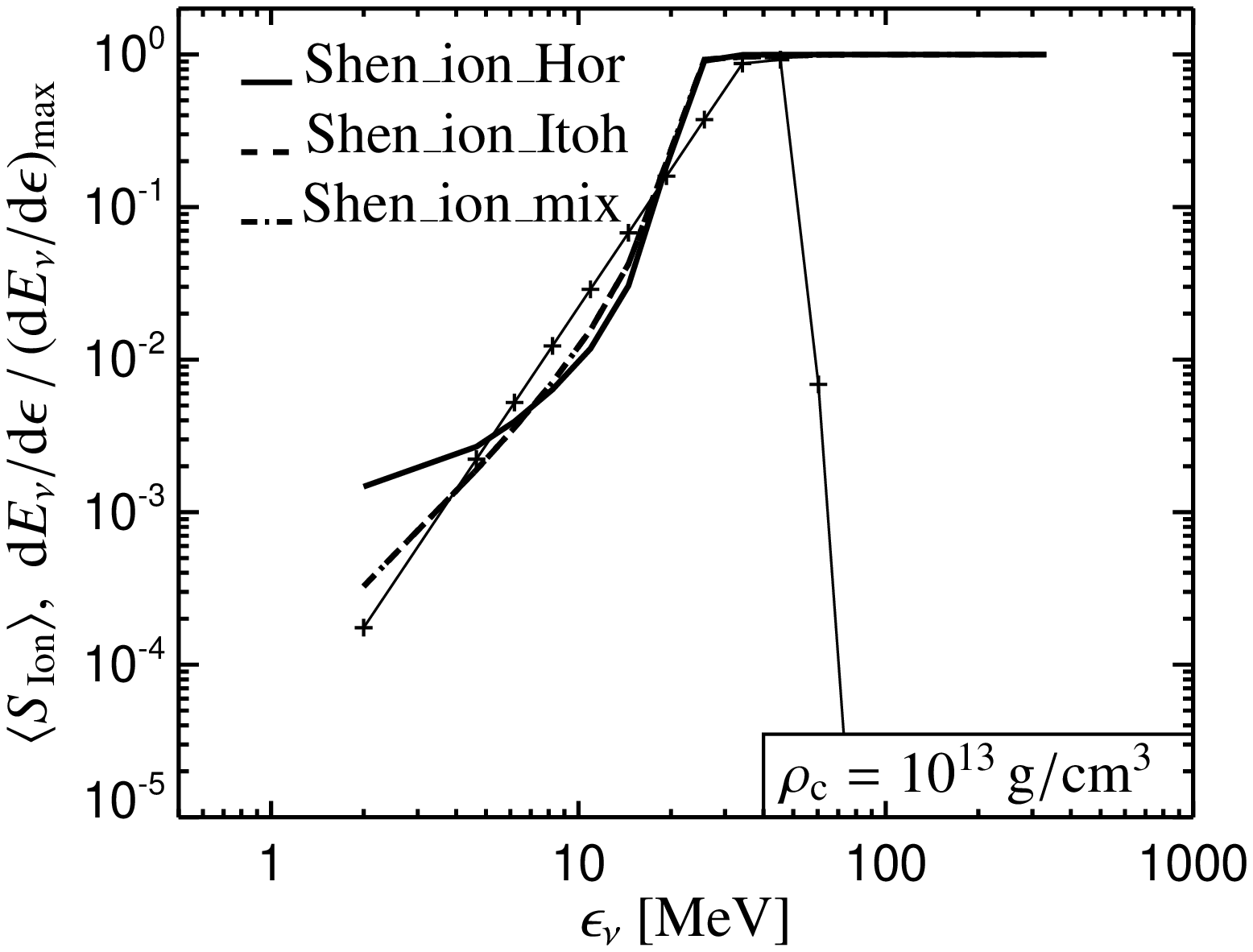}} &
\resizebox{0.5\linewidth}{!}{
\includegraphics{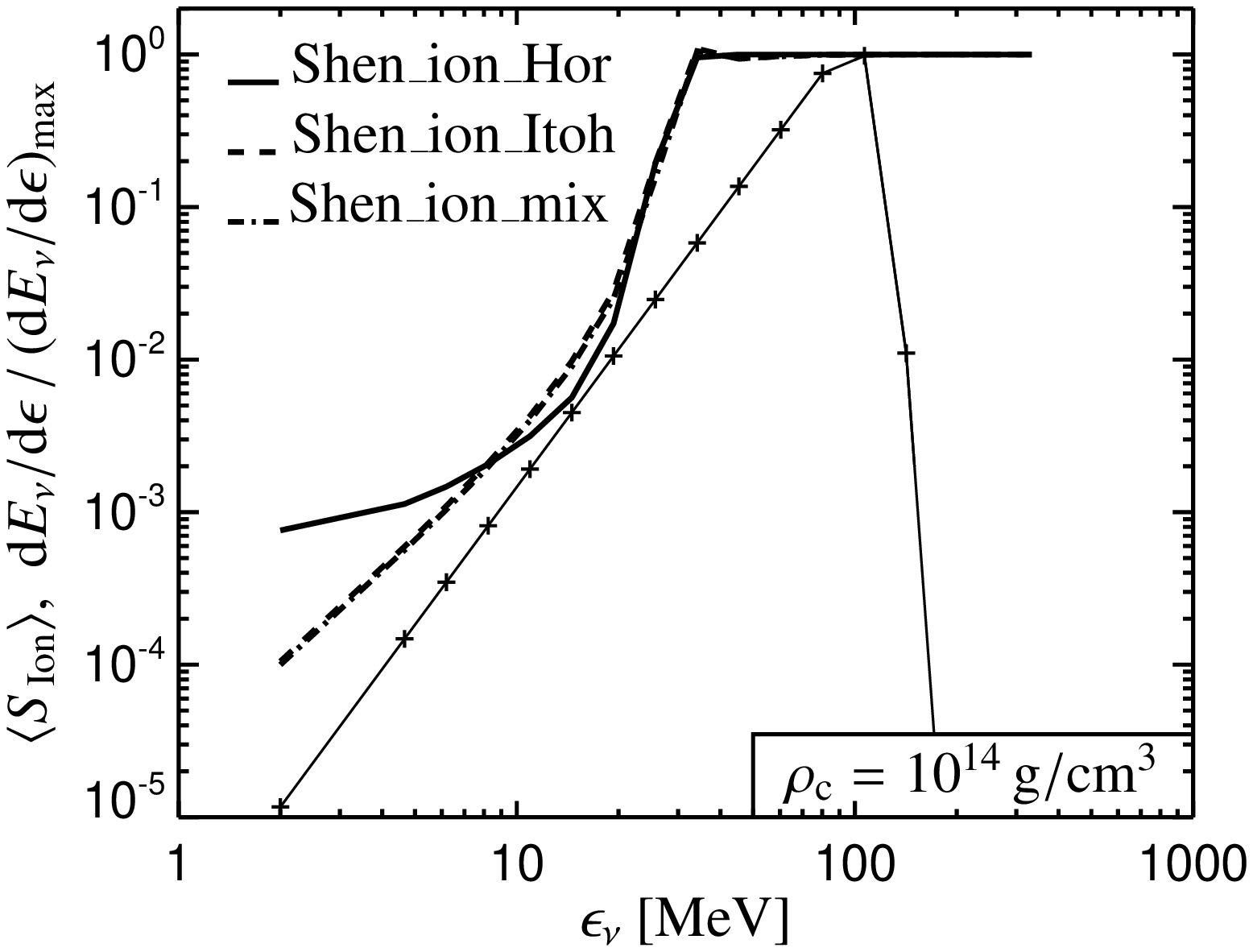}} \\
\bfseries{c} & \bfseries{d} \\
\end{tabular}
\caption{The cross section suppression factor
$\langle S_\mathrm{ion}\rangle$ for neutrino scattering off
heavy nuclei (bold lines) as function of
neutrino energy at the center of the iron core for densities of
(a) $10^{11}\,$g$\,$cm$^{-3}$, (b)     $10^{12}\,$g$\,$cm$^{-3}$,
(c) $10^{13}\,$g$\,$cm$^{-3}$, and (d) $10^{14}\,$g$\,$cm$^{-3}$.
The results are taken from collapse simulations with the 
\cite{shetok_1_98,shetok98} EoS. 
Also plotted (thin solid lines with crosses)
are the local spectra of the neutrino
energy density (normalized to the spectral maximum), which are
practically identical for all simulations.
}\label{fig:sfactor+spectra}
\end{figure*}

On their way out escaping neutrinos transfer a part of their
energy to electrons in collisions, thus heating the stellar
medium \citep[][]{bru86}. Since ion screening of neutrino-nucleus
scatterings reduces the transport opacity and therefore the 
effective optical depth for energy 
exchange with the stellar background mainly for low-energy
neutrinos, but hardly changes the downscattering probability
of high-energy neutrinos (Fig.~\ref{fig:opticaldepth}),
the larger loss of lepton number 
leads to an increase of the central entropy by about 
0.12$\,k_\mathrm{B}$ per nucleon (Fig.~\ref{fig:centrals}).
In Fig.~\ref{fig:opticaldepth} the optical depth $\tau$ at
the center is calculated as
\begin{equation}
\tau(\epsilon)=\int_0^\infty \mathrm{d}r\, 
{\left(\lambda_\mathrm{eff}(\epsilon,r)\right)^{-1}} \mbox{ ,}
\end{equation}
where ${\lambda_\mathrm{eff}(\epsilon,r)}$ is the effective mean
path for energy exchange, i.e., the average displacement between 
two reactions with energy exchange between neutrinos and target 
particles \citep[see][]{ryblight79}. It can be expressed as
\begin{equation}
{\lambda_\mathrm{eff}(\epsilon,r)}=\sqrt{\lambda_\mathrm{t}(\epsilon,r)\,
  \lambda_\mathrm{e}(\epsilon,r)} \mbox{ ,}
\end{equation}
when $\lambda_\mathrm{e}$ is the mean free path for reactions
with energy exchange, i.e.\ neutrino absorption and
neutrino-electron scattering in the present context,
and $\lambda_\mathrm{t}$ is the
total mean free path for momentum transfer (``transport mean
free path''), which includes all processes by which neutrinos
interact with the stellar fluid.

Figure~\ref{fig:sfactor+spectra} displays the ion screening 
factors $\langle S_\mathrm{ion}\rangle$
superimposed to the normalized neutrino energy spectra
$\mathrm{d}E_\nu/\mathrm{d}\epsilon_\nu$
at the stellar center for densities of 
$10^{11}\,$g$\,$cm$^{-3}$, $10^{12}\,$g$\,$cm$^{-3}$, 
$10^{13}\,$g$\,$cm$^{-3}$, and $10^{14}\,$g$\,$cm$^{-3}$.
Figure~\ref{fig:sfactor+spectra} once more demonstrates 
that the reduction of neutrino-nucleus scattering
mostly affects neutrinos at energies below the spectral
maximum for all plotted cases. In combination with
Fig.~\ref{fig:opticaldepth} it also shows that the trapping
conditions for the bulk of the neutrino spectrum are not
influenced strongly by ion screening. 
This was identified by \cite{brumez97}
as the reason why ion-ion correlations have no dramatic
effect on the core deleptonization. 

From Fig.~\ref{fig:comparesfactor} it is clear that
differences between the treatments of ion-ion correlations
by \cite{hor97} and \cite{Itoh2004} are largest for
$\xi \la 0.5$. The improvements by \cite{Itoh2004} are
therefore most important for the lowest neutrino energies 
in the energy window affected by ion screening. 
Since the phase space
available at such low energies is small, one cannot expect
large quantitative consequences for stellar core collapse.  
This is confirmed by 
Figs.~\ref{fig:centralye}--\ref{fig:shockformation} and
Fig.~\ref{fig:profiles}.
Itoh et al.'s (2004) description (in Model Shen\_ion\_Itoh) 
leads to values of $Y_e$, $Y_\mathrm{lep}$, and $s$ after
trapping which are essentially indistinguishable from those
obtained with Horowitz's (1997) formulae, consistent
with the insignificant differences
between Models Shen\_ion\_Hor and Shen\_ion\_Itoh seen
in the other plots. Note that the crossing of the 
$Y_e$-, $Y_\mathrm{lep}$- and $s$-profiles
for simulations with and without ion screening
at 0.45$\,M_\odot$ (Fig.~\ref{fig:profiles})
was also present in the results of \cite{brumez97}. 
Finally testing the sensitivity of the core collapse
evolution to the treatment of ion screening for the ionic 
mixture of free protons, $\alpha$ particles and a representative 
heavy nucleus, we also could not discover any differences of 
relevance. 

A more detailed analysis reveals the reasons for this
insensitivity, which are valid for both employed EoSs:
Below the neutrino trapping regime (i.e., for
$\rho \la 10^{12}\,$g$\,$cm$^{-3}$) even for low-energy
neutrinos ($\epsilon_\nu \approx 5\,$MeV) the parameter $\xi$
is larger than or around unity (except for neutrinos
interacting with $\alpha$ particles in an ionic mix when
$\xi_\alpha$ is computed from 
Eqs.~\ref{eqn:amixture}--\ref{eqn:ximixture}).
Moreover, $\Gamma \la 50$ holds
at the same time, implying that the ion-ion correlation
factors $\langle S_\mathrm{ion}(\xi,\Gamma) \rangle_\mathrm{Hor}$
and $\langle S_\mathrm{ion}(\xi,\Gamma) \rangle_\mathrm{Itoh}$
for neutrino scattering off heavy nuclei
are essentially the same (see Fig.~\ref{fig:comparesfactor}).
Only at densities above the trapping density the value of $\xi$
drops significantly below unity and $\Gamma$ exceeds 50,
causing visible (typically factors 2--3 for $\epsilon_\nu \la
5\,$MeV; Fig.~\ref{fig:sfactor+spectra}) differences in the 
ion-ion suppression factors $\langle S_\mathrm{ion} \rangle$.
At these densities, however, the
exact value of the neutrino-nucleus scattering cross section 
has no noticeable influence on the evolution of the core 
properties and on the neutrino transport.

Alpha particles in the inner core do not become
sufficiently abundant to cause mentionable differences. Their
indirect effect on ion-ion correlations of heavy nuclei
in an ion mixture by reducing $a_j$ (and thus $\xi_j$) 
and increasing $\Gamma_j$ (Sect.~\ref{sect:ionmixture})
for the heavier nuclei is essentially negligible, because their
contribution to the sum in Eq.~(\ref{eqn:electronsphere}) is 
diminished by their number density 
being multiplied with a factor $Z_\alpha/Z_j \ll 1$.
Moreover, $\alpha$ particles do not account
for a significant contribution to the total neutral-current
scattering opacity, because the opacity for coherent scattering
of neutrinos by nuclei $(Z,N,A)$ scales roughly with $N^2/A$  
and therefore is much smaller for $\alpha$ particles than for
heavy nuclei. For this reason the direct influence of $\alpha$ 
particles and thus of the suppression of their coherent
(elastic) scattering cross section for low neutrino 
energies is miniscule, despite the fact that $\xi_\alpha$ drops
below unity already at densities $\rho \la 10^{11}\,$g$\,$cm$^{-3}$
in a mixture with heavy nuclei.
(On the other hand, $\Gamma_\alpha$ turns out to be always 
less than unity and, following \cite{hor97}, is therefore 
set to $\Gamma_\alpha = 1$ for evaluating the angle-averaged
ion screening correction factor.)

\section{Conclusions}
\label{sect:conclusions}

In this paper we presented results from simulations of 
stellar core collapse with the aim to investigate the 
consequences of ion-ion correlations in neutrino-nucleus
scattering, comparing Itoh et al.'s (2004) improved 
description with an older one by \cite{hor97}.
We employed the EoS of \cite{shetok_1_98,shetok98} in addition to
Lattimer and Swesty's (1991) EoS and treated electron 
captures on heavy nuclei according to \cite{lan2003}, making nuclei dominant over protons in producing
neutrinos up to the density of the phase transition to 
nuclear matter.

Despite these differences in the input physics,
our models essentially confirmed the previous calculations by 
\cite{brumez97} who followed \cite{hor97} in their 
description of ion screening. 
Because ion screening is effective only in a low-energy
window where the available phase space is rather small,
the influence of ion-ion correlations during stellar
core collapse and on the formation of the supernova shock 
is moderate \citep[][]{brumez97}.

We found that the improvement by \cite{Itoh2004} does not
lead to any noticeable differences because it affects only
neutrinos of very low energies ($\la\,$5$\,$MeV) before 
trapping densities ($\rho \approx 10^{12}\,$g$\,$cm$^{-3}$) 
are reached. Differences at larger neutrino energies occur 
only at higher densities and thus do not affect the 
deleptonization and entropy evolution.
Effects due to the ionic mixture of free protons, $\alpha$
particles, and a representative heavy nucleus --- using the
linear mixing rule as suggested by \cite{Itoh2004} --- were found
to be negligibly small, too, mainly because the abundance 
of $\alpha$ particles in the inner regions of the 
collapsing stellar core is too low to affect the ion
screening of heavy nuclei indirectly 
(see Sect.~\ref{sect:ionmixture}). Alpha particles
do not contribute to the total opacity for 
elastic neutrino-nucleus scattering on a level where 
their ion screening (which becomes sizable only
when the mixture effects of Sect.~\ref{sect:ionmixture} 
are accounted for) might be relevant.

Improving the description of ion-ion correlations for
the complex mix of heavy nuclei with a large variety of
components, alpha particles, and free nucleons in
the supernova core, however, is desirable. Referring to 
multi-component calculations based on the Debye-H\"uckel
approximation in the limit of small momentum transfer,
\cite{saw05} argues that a range of $N/Z$ ratios in ionic
mixtures can protect against the strong ion screening
suppression of neutrino-nuclei scattering 
predicted by the effective averages
of one-component plasma parameters applied in the
current literature and in this work.
Moreover, the description of nearly free nucleons and nuclei
in NSE is expected to hold only up to a density
of about $10^{13}\,$g$\,$cm$^{-3}$. Above this density and
below the normal nuclear matter saturation density, a
pasta phase may develop with nucleons clustered in subtle
and complex shapes. Correlation effects for coherent
neutrino scattering can then not be treated within the single
heavy nucleus approximation \citep[e.g.,][]{hor04b,hor04a,watan04}.

\begin{acknowledgements}
We thank Naoki Itoh for providing us with a subroutine to 
compute his ion-ion correlation factor and for helpful
discussions about the treatment of ionic mixtures. We are
grateful to an anonymous referee for pointing out to us an
inaccurate use of an average ion sphere radius in the
case of ionic mixtures.
We also thank K.~Langanke, G.~Mart\'{\i}nez-Pinedo and 
J.M.~Sampaio for their table of electron capture rates on 
nuclei, which was calculated by employing a Saha-like NSE 
code for the abundances written by W.R.~Hix.
Support by the Sonderforschungsbereich 375
``Astro-Particle Physics'' of the Deutsche 
Forschungsgemeinschaft is acknowledged.
\end{acknowledgements}

\appendix

\bibliography{bibtex/aamnem,bibtex/lit}

\begin{thebibliography}{39}
\expandafter\ifx\csname natexlab\endcsname\relax\def\natexlab#1{#1}\fi

\bibitem[{{Bruenn}(1985)}]{bru85}
{Bruenn}, S.~W. 1985, \apjs, 58, 771

\bibitem[{{Bruenn}(1986)}]{bru86}
{Bruenn}, S.~W. 1986, \apj, 311, L69

\bibitem[{{Bruenn}(1989{\natexlab{a}})}]{bru89:xp}
{Bruenn}, S.~W. 1989{\natexlab{a}}, \apj, 340, 955

\bibitem[{{Bruenn}(1989{\natexlab{b}})}]{bru89:eos}
{Bruenn}, S.~W. 1989{\natexlab{b}}, \apj, 341, 385

\bibitem[{{Bruenn} \& {Mezzacappa}(1997)}]{brumez97}
{Bruenn}, S.~W. \& {Mezzacappa}, A. 1997, \prd, 56, 7529

\bibitem[{{Buras} {et~al.}(2003){Buras}, {Janka}, {Keil}, {Raffelt}, \&
  {Rampp}}]{burjan03:nunu}
{Buras}, R., {Janka}, H.-T., {Keil}, M.-T., {Raffelt}, G., \& {Rampp}, M. 2003,
  \apj, 587, 320

\bibitem[{{Burrows} \& {Sawyer}(1998)}]{bursaw98}
{Burrows}, A. \& {Sawyer}, R.~F. 1998, \prc, 58, 554

\bibitem[{{Burrows} \& {Sawyer}(1999)}]{bursaw99}
{Burrows}, A. \& {Sawyer}, R.~F. 1999, \prc, 59, 510

\bibitem[{{Carter} \& {Prakash}(2002)}]{carpra02}
{Carter}, G.~W. \& {Prakash}, M. 2002, Physics Letters B, 525, 249

\bibitem[{{Cernohorsky}(1994)}]{cer94}
{Cernohorsky}, J. 1994, \apj, 433, 247

\bibitem[{{Fryxell} {et~al.}(1989){Fryxell}, {M\"uller}, \&
  {Arnett}}]{frymue89}
{Fryxell}, B.~A., {M\"uller}, E., \& {Arnett}, W.~D. 1989, Hydrodynamics and
  Nuclear Burning, preprint MPA-449, Max Planck Institut f\"ur Astrophysik,
  Garching

\bibitem[{{Hannestad} \& {Raffelt}(1998)}]{hanraf98}
{Hannestad}, S. \& {Raffelt}, G. 1998, \apj, 507, 339

\bibitem[{{Heger} {et~al.}(2001){Heger}, {Woosley}, {Mart{\'{\i}}nez-Pinedo},
  \& {Langanke}}]{wooheg01}
{Heger}, A., {Woosley}, S.~E., {Mart{\'{\i}}nez-Pinedo}, G., \& {Langanke}, K.
  2001, \apj, 560, 307

\bibitem[{{Horowitz}(1997)}]{hor97}
{Horowitz}, C.~J. 1997, \prd, 55, 4577

\bibitem[{{Horowitz}(2002)}]{hor01}
{Horowitz}, C.~J. 2002, \prd, 65, 043001

\bibitem[{{Horowitz} {et~al.}(2004{\natexlab{a}}){Horowitz}, {P{\'
  e}rez-Garc{\'{\i}}a}, {Carriere}, {Berry}, \& {Piekarewicz}}]{hor04b}
{Horowitz}, C.~J., {P{\' e}rez-Garc{\'{\i}}a}, M.~A., {Carriere}, J., {Berry},
  D.~K., \& {Piekarewicz}, J. 2004{\natexlab{a}}, \prc, 70, 065806

\bibitem[{{Horowitz} {et~al.}(2004{\natexlab{b}}){Horowitz}, {P{\'
  e}rez-Garc{\'{\i}}a}, \& {Piekarewicz}}]{hor04a}
{Horowitz}, C.~J., {P{\' e}rez-Garc{\'{\i}}a}, M.~A., \& {Piekarewicz}, J.
  2004{\natexlab{b}}, \prc, 69, 045804

\bibitem[{{Itoh}(1975)}]{itoh75}
{Itoh}, N. 1975, Progress of Theoretical Physics, 54, 1580

\bibitem[{{Itoh} {et~al.}(2004){Itoh}, {Asahara}, {Tomizawa}, {Wanajo}, \&
  {Nozawa}}]{Itoh2004}
{Itoh}, N., {Asahara}, R., {Tomizawa}, N., {Wanajo}, S., \& {Nozawa}, S. 2004,
  \apj, 611, 1041

\bibitem[{{Itoh} {et~al.}(1983){Itoh}, {Mitake}, {Iyetomi}, \&
  {Ichimaru}}]{Itoh83}
{Itoh}, N., {Mitake}, S., {Iyetomi}, H., \& {Ichimaru}, S. 1983, \apj, 273, 774

\bibitem[{{Itoh} {et~al.}(1979){Itoh}, {Totsuji}, {Ichimaru}, \&
  {Dewitt}}]{itoh79}
{Itoh}, N., {Totsuji}, H., {Ichimaru}, S., \& {Dewitt}, H.~E. 1979, \apj, 234,
  1079

\bibitem[{{Langanke} {et~al.}(2003){Langanke}, {Mart{\'{\i}}nez-Pinedo},
  {Sampaio}, {Dean}, {Hix}, {Messer}, {Mezzacappa}, {Liebend{\" o}rfer},
  {Janka}, \& {Rampp}}]{lan2003}
{Langanke}, K., {Mart{\'{\i}}nez-Pinedo}, G., {Sampaio}, J.~M., {et~al.} 2003,
  Physical Review Letters, 90, 241102

\bibitem[{{Lattimer} {et~al.}(1985){Lattimer}, {Pethick}, {Ravenhall}, \&
  {Lamb}}]{latpet85}
{Lattimer}, J.~M., {Pethick}, C.~J., {Ravenhall}, D.~G., \& {Lamb}, D.~Q. 1985,
  Nucl.~Phys.~A, 432, 646

\bibitem[{{Lattimer} \& {Swesty}(1991)}]{latswe91}
{Lattimer}, J.~M. \& {Swesty}, F.~D. 1991, Nucl.~Phys.~A, 535, 331

\bibitem[{{Liebend{\"o}erfer} {et~al.}(2005){Liebend{\"o}erfer}, {Rampp},
  {Janka}, \& {Mezzacappa}}]{lieram05}
{Liebend{\"o}erfer}, M., {Rampp}, M., {Janka}, H.-T., \& {Mezzacappa}, A. 2005,
  \apj, 620, 840

\bibitem[{{Marek} {et~al.}(2005){Marek}, {Dimmelmeier}, {Janka}, {M\"uller}, \&
  {Buras}}]{mardim05}
{Marek}, A., {Dimmelmeier}, H., {Janka}, H.-T., {M\"uller}, E., \& {Buras}, R.
  2005, astro-ph/0502161,\, \aap\, in press

\bibitem[{{Mart{\'{\i}}nez-Pinedo} {et~al.}(2005){Mart{\'{\i}}nez-Pinedo},
  {Langanke}, {Sampaio}, {Dean}, {Hix}, {Messer}, {Mezzacappa},
  {Liebend\"orfer}, {Janka}, \& {Rampp}}]{martinezpinedo05}
{Mart{\'{\i}}nez-Pinedo}, G., {Langanke}, K., {Sampaio}, J.~M., {et~al.} 2005,
  in Cosmic Explosions, IAU Colloquium 192 (Springer, Berlin), 321

\bibitem[{{Mezzacappa} \& {Bruenn}(1993{\natexlab{a}})}]{mezbru93:nes}
{Mezzacappa}, A. \& {Bruenn}, S.~W. 1993{\natexlab{a}}, \apj, 410, 740

\bibitem[{{Mezzacappa} \& {Bruenn}(1993{\natexlab{b}})}]{mezbru93:coll}
{Mezzacappa}, A. \& {Bruenn}, S.~W. 1993{\natexlab{b}}, \apj, 405, 637

\bibitem[{{Pons} {et~al.}(1998){Pons}, {Miralles}, \& {Ibanez}}]{ponmir98}
{Pons}, J.~A., {Miralles}, J.~A., \& {Ibanez}, J.~M.~A. 1998, \aaps, 129, 343

\bibitem[{{Rampp} \& {Janka}(2002)}]{ramjan02}
{Rampp}, M. \& {Janka}, H.-T. 2002, \aap, 396, 361

\bibitem[{{Reddy} {et~al.}(1999){Reddy}, {Prakash}, {Lattimer}, \&
  {Pons}}]{redpra99}
{Reddy}, S., {Prakash}, M., {Lattimer}, J.~M., \& {Pons}, J.~A. 1999, \prc, 59,
  2888

\bibitem[{{Rybicki} \& {Lightman}(1979)}]{ryblight79}
{Rybicki}, G.~B. \& {Lightman}, A.~P. 1979, {Radiative Processes in
  Astrophysics} (John Wiley and Sons, New York)

\bibitem[{{Sawyer}(2005)}]{saw05}
{Sawyer}, R.~F. 2005, astro-ph/0505520

\bibitem[{{Shen} {et~al.}(1998{\natexlab{a}}){Shen}, {Toki}, {Oyamatsu}, \&
  {Sumiyoshi}}]{shetok_1_98}
{Shen}, H., {Toki}, H., {Oyamatsu}, K., \& {Sumiyoshi}, K. 1998{\natexlab{a}},
  Nucl.~Phys.~A, 637, 435

\bibitem[{{Shen} {et~al.}(1998{\natexlab{b}}){Shen}, {Toki}, {Oyamatsu}, \&
  {Sumiyoshi}}]{shetok98}
{Shen}, H., {Toki}, H., {Oyamatsu}, K., \& {Sumiyoshi}, K. 1998{\natexlab{b}},
  Progress of Theoretical Physics, 100, 1013

\bibitem[{{Swesty} {et~al.}(1994){Swesty}, {Lattimer}, \& {Myra}}]{swe94}
{Swesty}, F.~D., {Lattimer}, J.~M., \& {Myra}, E.~S. 1994, \apj, 425, 195

\bibitem[{{Thompson} {et~al.}(2003){Thompson}, {Burrows}, \&
  {Pinto}}]{thobur03}
{Thompson}, T.~A., {Burrows}, A., \& {Pinto}, P.~A. 2003, \apj, 592, 434

\bibitem[{{Watanabe} {et~al.}(2004){Watanabe}, {Sato}, {Yasuoka}, \&
  {Ebisuzaki}}]{watan04}
{Watanabe}, G., {Sato}, K., {Yasuoka}, K., \& {Ebisuzaki}, T. 2004, \prc, 69,
  055805

\end{thebibliography}
\bibliographystyle{aa}

\end{document}